\pdfoutput=1

\documentclass[12pt,a4paper]{iopart}

\usepackage{iopams}
\usepackage{setstack}
\usepackage{graphicx}

\usepackage{epstopdf}
\usepackage{hyperref}

\newcommand{\sci}[2]{\ensuremath{\times 10^{#1}~\mathrm{#2}}}

\newcommand{\evat}[3][]{\ensuremath{\left.#2\right|^{#1}_{#3}}}
\newcommand{\deriv}[3][]{\ensuremath{\frac{\rmd^{#1} #2}{{\rmd #3}^{#1}}}}
\newcommand{\pderiv}[3][]{\ensuremath{\frac{\partial^{#1} #2}{{\partial #3}^{#1}}}}
\newcommand{\derivat}[4][]{\ensuremath{\evat{\deriv[#1]{#2}{#3}}{#4}}}

\newcommand{\hfrac}[2]{\ensuremath{\left.#1\middle/#2\right.}}

\newcommand{\fr}[1]{\ensuremath{\frac{1}{#1}}}

\newcommand{\ahalf}{\ensuremath{\fr{2}}}
\newcommand{\athird}{\ensuremath{\fr{3}}}

\newcommand{\cte}{\ensuremath{\rm{const}}}

\newcommand{\sandwich}[3]{\ensuremath{\langle #1 | #2 | #3 \rangle}}

\newcommand{\rs}{r^*}
\newcommand{\rc}{r_0}
\newcommand{\rst}{r_{\rm s}}
\newcommand{\ub}{\bar{u}}
\newcommand{\vb}{\bar{v}}
\newcommand{\dt}{\Delta t_0}
\newcommand{\dtw}{\widetilde{\Delta t}_0}
\newcommand{\UH}{U_{\rm H}}
\newcommand{\uH}{u_{\rm H}}

\newcommand{\pv}{p_{\rm vs}}
\newcommand{\ps}{p_{\rm so}}
\newcommand{\pf}{p_{\rm ffio}}
\newcommand{\pr}{p_{\rm ffro}}

\begin{document}

\title{Hawking radiation as perceived by different observers}

\author{L~C~Barbado$^1$, C~Barcel\'o$^1$ and L~J~Garay$^{2, 3}$}

\address{$^1$ Instituto de Astrof\'{\i}sica de Andaluc\'{\i}a (CSIC), Glorieta de la Astronom\'{\i}a, 18008 Granada, Spain}
\address{$^2$ Departamento de F\'{\i}sica Te\'orica II, Universidad Complutense de Madrid, 28040 Madrid, Spain}
\address{$^3$ Instituto de Estructura de la Materia (CSIC), Serrano 121, 28006 Madrid, Spain}

\eads{\mailto{luiscb@iaa.es}, \mailto{carlos@iaa.es}, \mailto{luis.garay@fis.ucm.es}}

\begin{abstract}

We use a method recently introduced in Barcel\'o {\it et al,} 10.1103/PhysRevD.83.041501, to 
analyse  Hawking radiation in a Schwarzschild black hole as perceived by 
different observers in the system. The method is based on the introduction of an 
``effective temperature'' function that varies with time. First we introduce a 
non-stationary vacuum state for a quantum scalar field, which interpolates between 
the Boulware vacuum state at early times and the Unruh vacuum 
state at late times. In this way we mimic the process of switching on Hawking radiation in 
realistic collapse scenarios. Then, we analyse this vacuum state from the perspective of 
static observers at different radial positions, observers undergoing a 
free-fall trajectory from infinity, and observers standing at rest at a radial
distance and then released to fall freely towards the horizon. The physical image that 
emerges from these analyses is rather rich and compelling. Among many other results, we find
 that generic freely-falling observes do not perceive vacuum when 
crossing the horizon, but an effective temperature a few times larger than the one 
that they perceived when started to free-fall. We explain this phenomenon as due to a 
diverging Doppler effect at horizon crossing.
\end{abstract}


\submitto{\CQG}


%--------------------------------------------------------------
\section{Introduction}
\label{Sec:introduccion}
%--------------------------------------------------------------

Arguably, the discovery that black holes should evaporate by emitting a Planckian spectrum of particles, is the most important result of combining general relativistic with quantum mechanical notions~\cite{hawking1, hawking2}. As it is by now well known, Hawking radiation is a kinematic effect (that is, does not directly depend on Einstein's equations), and so, much more general than its incarnation in General Relativity. In any phenomenon in nature describable in terms of quantum excitations propagating in a Lorentzian geometry with black-hole-like properties, one expects to have Hawking particle (or quasi-particle) production. In~\cite{lrr}, for example, you can find a catalogue of systems and situations outside the General Relativity realm in which Hawking radiation is expected to appear.
Here, by black-hole-like properties we mean either the presence of trapping horizons or just the asymptotic approach towards configurations with horizons~\cite{Barcelo:2010pj, trapping, grove}.

When the Lorentzian geometry acquires the form of a stationary black hole, the temperature of Hawking radiation becomes constant and proportional to the surface gravity of the black hole at the horizon~\cite{hawking2}. Moreover, this temperature only depends on this constant, leaving no trace of the history of the black hole geometry formation or preparation~\cite{hawking2, parentani}. In the General Relativity realm that we shall adopt in this paper, this surface gravity and so Hawking temperature, happens to be inversely proportional to the mass of the black hole.

When investigating Hawking radiation, it is customary to work on a simple static background geometry, namely the Schwarzschild solution, neglecting any back-reaction on the geometry. In order to have Hawking radiation at infinity, it is necessary to set the quantum field in a particular quantum state: the Unruh vacuum state~\cite{unruh, fabbri}. This state is commonly described as being a vacuum state for observers freely falling at the event horizon of the black hole. It is a stationary state, describing a black hole which has always been and will ever be emitting radiation at Hawking temperature at all times. Therefore, it does not capture the process of the black hole formation in General Relativity, in which Hawking radiation switches on in the last stages of collapse. Instead, in this paper we are going to analyse the characteristics of a different vacuum state, which might be called the collapse vacuum. This vacuum is appropriately chosen so as to mimic the process of switching on of Hawking radiation. Initially, there will be no radiation at infinity, but at some particular time, some radiation will start to be noticeable. After some rather quick transient regime, this radiation will become thermal with Hawking temperature (as in this paper we are not going to consider the effects of the evaporation, it will not be necessary to distinguish between the terms thermal and Planckian, and we will use them indistinctively). Our collapse vacuum interpolates between 
the Boulware vacuum state~\cite{boulware} at early times and the Unruh vacuum state at late times.

It is well known that ``presence of particles'' is an observer-dependent notion~\cite{unruh, birrell-davies}. In this paper we will be specifically interested in analysing how the collapse vacuum state is perceived by different observers. In particular, we are going to consider (i) static observers at different radial positions of spacetime, (ii) observers undergoing a free-fall trajectory from infinity, passing through a fixed radial position at different times, and (iii) observers standing at rest at a radial distance until a moment at which they are released to fall freely towards the hole. A discussion about the different perceptions of the radiation from a black hole in the Unruh state by freely-falling and stationary observers can be found in~\cite{greenwood}.

In order to analyse this particle perception, we could use Bogoliubov coefficients~\cite{birrell-davies}, but these, except for a few particular simple cases, are difficult to calculate. Instead, we could use the functional Schrodinger formalism as in~\cite{greenwood}, but it appears difficult to manipulate for non-stationary vacuum states. In this paper we will adopt a different strategy. Following~\cite{Barcelo:2010xk}, we shall define and calculate a time dependent effective temperature for each observer along its world line. Then, abusing of the language we will describe the perception of the observer as experiencing a thermal radiation wind with a time dependent temperature. This effective temperature will only be a real temperature in situations in which an additional adiabatic condition is satisfied. We will also calculate when this condition is satisfied and when it is not. Even in the cases in which the adiabatic condition is violated, and therefore the radiation spectrum differs from the Planckian shape, it will be argued that the calculated effective temperature is still useful as an estimator of the amount of particles encountered by the observer at each instant of time. But it is important to remark that in these cases our method cannot give information about the characteristics of the spectrum. In any case one should keep in mind that a precise calculation of the particle perception would require the construction of a particle detector, taking into account its spatial and temporal resolutions. This is, however, beyond the scope of this paper. Nonetheless, we will show that the physical image that emerges from our description is rather rich and compelling. For instance, it will allow us to clearly see that freely-falling observers at the horizon will not perceive vacuum unless (a) they closely follow the collapse or (b) they have instantaneously zero radial velocity at the horizon. We will also show that this result is not an artifact of the chosen vacuum, but that it is precisely a characteristic of the final Unruh state.

The paper is structured as follows. In \sref{preliminaries}, we will start recalling the basic background ingredients needed for our analysis. In \sref{vacuum.choice}, we will describe and fix the vacuum state for the quantum radiation field of our problem. Then, in \sref{radiation.observers}, after describing what we mean by effective temperature and adiabatic control functions, we will study in detail the radiation perception associated with the already enumerated different observers. We will devote \sref{Sec:interpretation} to the explanation of one of our most important results: the non-zero perception of radiation by freely-falling observers at the horizon, against the opposite naive expectation. After a brief comparison of the results with those that one would obtain in a pure Unruh state (\sref{Sec:Unruh}), we will finally summarize and conclude.

\section{Preliminaries}\label{preliminaries}

\subsection{Geometrical set up}

We shall restrict ourselves to the analysis of radiation emission in a Schwarzschild black hole. In units where $G=c=1$, the Schwarzschild metric reads
\begin{equation}
\rmd s^2 = -\left( 1-\frac{2m}{r} \right) \rmd t^2 + \left( 1-\frac{2m}{r} \right)^{-1} \rmd r^2 + r^2 \rmd \Omega_2^2,
\label{metric}
\end{equation}
with $m$ being the mass of the black hole. Moreover, we will only use the $(t, r)$ sector of the geometry. In double null coordinates, the corresponding  $(1+1)$-metric can be written as
\begin{equation}
\rmd s^2 = -\left( 1-\frac{2m}{r} \right) \rmd \ub \rmd \vb.
\label{metric.null}
\end{equation}
Here,
\begin{equation}
\ub := t-\rs, \qquad \vb := t+\rs,
\label{rays}
\end{equation}
with $\rs$ being the ``tortoise'' coordinate
\begin{equation}
\rs := r + 2m \ln \left( \frac{r}{2m} - 1 \right).
\label{tortoise}
\end{equation}
%

%--------------------------------------------------------------
\subsection{Quantum field theory}
\label{quantum.field.theory}
%--------------------------------------------------------------

We are going to analyse the behaviour of a quantum field in this spacetime. For simplicity, we will only consider a Klein-Gordon massless scalar field (analyses with other fields would yield qualitatively similar results). In this paper, we will only consider spherically symmetric configurations and ignore any backscattering of the scalar field in the geometry, so that our analysis will be equivalent to that in a $(1+1)$-dimensional conformally invariant theory (here on, we will stick to a pure $1+1$ analysis). Now, the $1+1$ Klein-Gordon equation in null coordinates reads
\begin{equation}
\pderiv{}{\ub}\pderiv{}{\vb} \phi (\ub, \vb) = 0,
\label{KG.eq}
\end{equation}
so that its general solution can be written as
\begin{equation}
\phi = f(\ub) + g(\vb).
\label{general.solution}
\end{equation}
Owing to the conformal invariance of the $1+1$ Klein-Gordon equation, any relabelling $\ub = \pv (U)$, $\vb = q_{\rm vs} (V)$ (the subscript `vs' in these functions stands for \emph{vacuum state}; the reason for this notation will be understood later on) can be associated with a mode decomposition of the quantum field of the form
\begin{equation}
\hat \phi = \int_0^\infty \rmd \omega' \left[ \hat a_{\omega'}^U \phi_{\omega'}^U + \hat a_{\omega'}^V \phi_{\omega'}^V + 
\hat a_{\omega'}^{U\dagger} (\phi_{\omega'}^U)^* + \hat a_{\omega'}^{V\dagger} (\phi_{\omega'}^V)^* \right],
\label{mode.decomposition}
\end{equation}
where the normalized modes in this expression are
\begin{equation}
\phi_{\omega'}^U ={1 \over \sqrt{4\pi \omega'}} \rme^{-\rmi \omega' U}, \qquad
\phi_{\omega'}^V ={1 \over \sqrt{4\pi \omega'}} \rme^{-\rmi \omega' V}.
\label{modes}
\end{equation}
This can easily be seen by using the following form of the Klein-Gordon scalar product
\begin{equation}
\langle \phi_1, \phi_2 \rangle := - \rmi\left( - \int  \rmd U \phi_1  \stackrel{\leftrightarrow}{\partial}_U \phi_2^* + \int \rmd V \phi_1  \stackrel{\leftrightarrow}{\partial}_V \phi_2^* \right).
\label{scalar.product}
\end{equation}
Once the functions $\ub = \pv (U)$, $\vb = q_{\rm vs} (V)$ have been defined, one can select a vacuum state $|0'\rangle$ by requiring that it satisfies
\begin{equation}
\hat a_{\omega'}^U |0' \rangle = 0, \qquad \hat a_{\omega'}^V |0' \rangle = 0.
\label{vacuum.definition}
\end{equation}
For any other choice $\ub = p_{\rm ob} (u)$, $\vb = q_{\rm ob} (v)$, we can perform an analogue mode decomposition and also define a vacuum state,
\begin{equation}
\hat a_\omega^u |0 \rangle = 0, \qquad \hat a_\omega^v | 0 \rangle = 0.
\label{vacuum.definition2}
\end{equation}
In general, the vacuum states $|0' \rangle$ and $|0 \rangle$ will be different. Thus, an observer for which the vacuum is $|0 \rangle$ might detect particles corresponding to the vacuum $|0' \rangle$.

%--------------------------------------------------------------
\subsection{Bogoliubov coefficients}\label{Sec:Bogoliubov}
%--------------------------------------------------------------

The particle content of the vacuum $|0' \rangle$ as seen by observers with an unprimed vacuum notion can be calculated through Bogoliubov coefficients. In 
this paper we will only worry about outgoing radiation, so we will forget about the 
$\phi_\omega^v$ modes, which correspond to rays going into the black hole. 
For each frequency $\omega$, we have (see~\cite{birrell-davies})
\begin{equation}
\sandwich{0'}{ N_\omega^u }{0'} = \sandwich{0'}{ a_\omega^{u\dagger} a_\omega^u }{0'} = \int \rmd \omega' \left| \beta_{\omega \omega'} \right|^2,
\label{number.particles}
\end{equation}
where $\beta_{\omega \omega'}$ is the Bogoliubov coefficient defined by the scalar product
\begin{equation}
\beta_{\omega \omega'} := - \langle \phi_\omega^u, (\phi_{\omega'}^U)^* \rangle,
\label{beta}
\end{equation}
with $\phi_\omega^u$ defined analogously to~\eref{modes}. Plugging the 
expressions for these modes into~\eref{scalar.product}, one arrives 
at
\begin{equation}
\beta_{\omega \omega'} = \fr{4 \pi \sqrt{\omega \omega'}} \left( \omega \int \rmd u \rme^{-\rmi (\omega' U + \omega u)} - \omega' \int \rmd U \rme^{-\rmi (\omega' U + \omega u)} \right),
\label{scalar.prod.integral.prev}
\end{equation}
or, integrating by parts the second term, at
\begin{equation}
\beta_{\omega \omega'} = \fr{2 \pi} \sqrt{\frac{\omega}{\omega'}} \int \rmd u \rme^{-\rmi \omega' U(u)}\rme^{-\rmi \omega u},
\label{scalar.prod.integral}
\end{equation}
(where we have eliminated an irrelevant boundary term). Therefore, to determine the particle content, one only needs to know the relation $U(u)$.

%--------------------------------------------------------------
\section{Vacuum state choice}\label{vacuum.choice}
%--------------------------------------------------------------

We are interested in analysing how Hawking radiation is perceived by different observers in spacetime. In order to have Hawking radiation in the first place, we need to select an appropriate vacuum state. Instead of making the obvious choice of the Unruh vacuum, we shall choose as vacuum state a non-stationary state which somehow interpolates between the Boulware state (with no radiation at both past and future asymptotic infinities) at early times and the Unruh state at late times. This dynamical (or non-stationary) state in a Schwarzschild static background will mimic what happens in a collapse process, in which the black hole is generated from an initially (almost) Minkowskian spacetime. As opposed to what happens in the Unruh vacuum, in this situation Hawking radiation will not be present at early times, but will be switched on at a particular ignition time (this would correspond to the time of the formation of the  black hole).

To impose this vacuum state to the quantum field, we will consider a time-like 
geodesic observer falling into the Schwarzschild black hole from infinity. We will then 
calculate how this observer would label in a natural way the different outgoing 
($\ub=\cte$) rays that he encounters in its way towards the horizon. This
 new label $U = \pv^{-1} (\ub)$ will then be used to define the 
vacuum 
state via a positive-frequency mode expansion, as explained in 
\sref{quantum.field.theory} [see~\eref{mode.decomposition}].

So let us first calculate the radial time-like geodesic trajectories by solving the geodesic equation
\begin{equation}
 \deriv[2]{r}{\tau} = - \Gamma^r_{r r} \left( \deriv{r}{\tau} \right)^2 - 
\Gamma^r_{t t} \left( \deriv{t}{\tau} \right)^2.
\label{geodesic.ref}
\end{equation}
With this aim we will take into account the relation
\begin{equation}
  \left(\deriv{t}{\tau}\right)^2 = \left( 1-\frac{2m}{r} \right)^{-1} \left[1 + \left( 1-\frac{2m}{r} \right)^{-1} \left(\deriv{r}{\tau}\right)^2 \right]
\label{metric.deriv}
\end{equation}
that follows from the spacetime interval (equivalently, we could use directly a variational principle for the spacetime interval). Here, $\tau$ is the proper time of the trajectory, and $\Gamma^r_{r r}$ and $\Gamma^r_{t t}$ are the only non-vanishing Christoffel symbols relevant for the calculation of radial trajectories,
\begin{equation}
\Gamma^r_{r r} = -\frac{m}{r^2} \left( 1-\frac{2m}{r} \right)^{-1}, \qquad \Gamma^r_{t t} = \frac{m}{r^2} \left( 1-\frac{2m}{r} \right).
\label{chris}
\end{equation}
Replacing them in  the radial geodesic equation yields
\begin{equation}
\deriv[2]{r}{\tau} = -\frac{m}{r^2},
\label{Newton}
\end{equation}
which has a form resembling Newton's gravitational force. Integrating once, we obtain
\begin{eqnarray}
\deriv{r}{\tau} = -\sqrt{\frac{2m}{r}}\left(1  -\frac{r}{\rc} \right)^{1/2},
\label{r.tau.first.integral}
\\
\deriv{t}{\tau} = \left(1- {2m \over r}\right)^{-1} \left(1  -\frac{2m}{\rc} \right)^{1/2},
\label{t.tau.first.integral}
\end{eqnarray}
in which we use the radius $\rc$ at which the velocity vanishes  as an  integration constant. For the time being, we are interested in describing a free-fall trajectory starting at radial infinity, that is, with $\rc \to \infty$. Then, integrating the radial equation we obtain
\begin{equation}
r = \left[ \frac{3 \sqrt{2m}}{2} (\tau_0-\tau) \right]^{2/3},
\label{r.of.tau}
\end{equation}
with $\tau_0$ being  an integration constant. Note that $\tau$ runs from $-\infty$ 
to $\tau_0$, at which time the trajectory reaches $r=0$. Before that, at
$\tau_0 - 4m/3$ the event horizon is crossed. We can also find the behaviour of the
 time coordinate $t$ with respect to $\tau$. Indeed, dividing equations 
\eref{r.tau.first.integral} and \eref{t.tau.first.integral} (with $\rc \to \infty$), we 
obtain
\begin{equation}
\deriv{r}{t} = - \sqrt{\frac{2m}{r}} \left( 1-\frac{2m}{r} \right),
\label{dr.of.t}
\end{equation}
which can be integrated to yield
\begin{equation}
t= t_0 -4m \left[ \sqrt{\frac{r}{2m}} + \athird \left( \frac{r}{2m} \right)^{3/2} + \ahalf \ln \left( \frac{\sqrt{r/(2m)}-1}{\sqrt{r/(2m)}+1} \right) \right],
\label{t.of.r}
\end{equation}
where $t_0$ is an integration constant. Finally, replacing the
relation~\eref{r.of.tau} for $r(\tau)$ in this expression, we obtain
\begin{equation}
\fl t= t_0 - 4m \left\{ \left[\frac{3 (\tau_0-\tau)}{4 m} \right]^{1/3} + \frac{\tau_0-\tau}{4m} + \ahalf \ln \left[ \frac{(3 (\tau_0-\tau) / 4m)^{1/3} - 1}{(3 (\tau_0-\tau) / 4m)^{1/3} + 1} \right] \right\}.
\label{t.of.tau}
\end{equation}
Now, having the pair $(t, r)$ as a function of $\tau$ in~\eref{r.of.tau} and~\eref{t.of.tau}, we can also write the quantity $\ub$ in~\eref{rays} as a function of $\tau$,
\begin{eqnarray}
\fl \ub =t_0- 4m \left\{ \left[ \frac{3 (\tau_0-\tau)}{4 m} \right]^{1/3} + \ahalf \left[ \frac{3 (\tau_0-\tau)}{4 m} \right]^{2/3} + \frac{\tau_0-\tau}{4m}
\right.
\nonumber\\
\left. + \ln \left[ \left(\frac{3 (\tau_0-\tau)}{4m}\right)^{1/3}  - 1 \right] \right\}.
\label{ub.of.tau}
\end{eqnarray}

An observer following this world line can naturally use its proper time $\tau$ to label the different rays he encounters in its way towards the horizon, namely, he can make the assignment of the label $U(\ub)=\tau$, given by the inverse of \eref{ub.of.tau}, to the ray $\ub=\cte$ that hits the observer at proper time $\tau$. Then, defining $\UH := \tau_0-4m/3$ and choosing the two constants $t_0=\tau_0$ so that the two labels $U$ and $\ub$ are synchronized at $\ub \to -\infty$, in the sense that $U = U(\ub \to -\infty) \to \ub$, the final relation  between both labels is
\begin{eqnarray}
\fl \ub = \pv(U) := \UH + \frac{4m}{3} - 4m \left\{ \left[ {3 \over 4m}(\UH-U) +1 \right]^{1/3} + \ahalf \left[ {3 \over 4m}(\UH-U) +1 \right]^{2/3}
\right.
\nonumber \\ 
+ \left.  \athird \left[ {3 \over 4m}(\UH-U) +1 \right] + \ln \left[ \left({3 \over 4m}(\UH-U) +1\right)^{1/3}  - 1 \right] \right\}.
\label{ub.of.U}
\end{eqnarray}
Note that, while the range of the original null label is $\ub \in 
(-\infty,+\infty)$, the range of the new null label is $U \in (-\infty,\UH)$, where 
$\UH$ marks the moment at which the falling observer crosses the event horizon. 
As explained before, associated with this choice  $U=\pv^{-1} (\ub)$, we have a 
vacuum state, and it is this vacuum state which will be the subject of our analysis.

%--------------------------------------------------------------
\section{Radiation perception for different observers}\label{radiation.observers}
%--------------------------------------------------------------

In this section, we will investigate what is the particle content of the vacuum state 
defined above  for different types of observers, for which the notion of 
vacuum might or might not coincide with this one. More specifically, we 
are 
going to  define yet another label $u = p^{-1}_{\rm ob} (\ub)$ associated, in 
each case, with a specific type of observer. Once we have characterized the 
observer by the null-ray labelling function, we will find the function 
$U=\pv^{-1}[p_{\rm ob} (u) ] =: U(u)$ which will inform us about the (observer-dependent) particle content of the chosen vacuum state.

\subsection{Variable temperature-like estimator}

As an alternative to the calculation of the exact Bogoliubov coefficients using~\eref{scalar.prod.integral}, we can estimate the amount of particle content by calculating the function
\begin{equation}
\kappa (u) := -\deriv[2]{U}{u} \left( \deriv{U}{u} \right)^{-1}.
\label{kappa.def}
\end{equation}
As described thoroughly in~\cite{Barcelo:2010pj, Barcelo:2010xk}, when this function is constant $\kappa(u) \simeq \kappa (u_*)=:\kappa_*$ over a sufficiently large interval around a given $u_*$, one can assure that during this same interval the system is producing a Hawking flux of particles with a temperature
\begin{equation}
k_{\rm B} T = {\kappa_* \over 2\pi}.
\label{hawking.temperature}
\end{equation}
If the variation of $\kappa(u)$ is slow, then one can describe the system as a 
thermal emitter with a slowly varying temperature. Whether $\kappa(u)$ varies 
slowly or not in the surroundings of $u^*$, is controlled by an adiabatic condition 
which, under mild technical assumptions~\cite{Barcelo:2010xk}, reads
\begin{equation}
\epsilon_* := {1 \over \kappa_*^2} \left| \evat{\deriv{\kappa}{u}}{u^*} \right| \ll 1.
\label{adiabatic.condition.alt}
\end{equation}

However, even in the case in which the adiabatic condition is not satisfied, the 
existence of a non-zero $\kappa(u)$ is an indication that there is particle
 emission (in this case the Bogoliubov $\beta$ coefficients will not be zero). The 
spectrum of this particle content does not follow a precise Planckian profile anymore,
 but still we will take $\kappa(u)$ as an estimator of the amount of 
particles perceived by the observer. In this work, for short we will refer to $\kappa(u)$ [or more precisely to $\kappa (u)/ (2\pi)$] as a 
variable temperature, although strictly speaking this term is correct only in the 
intervals in which the adiabatic condition is satisfied. We will call 
\begin{equation}
\epsilon (u) := {1 \over \kappa^2} \left| \deriv{\kappa}{u} \right|
\label{adiabatic.condition}
\end{equation}
the adiabatic control function

%-----------------------------------------------------------------------
\subsection{Static observer at a fixed radius}
\label{static.observer}
%-----------------------------------------------------------------------

Let us start by analysing the particle content as seen by a static observer sitting at a radius $\rst$. In order to find a new labelling $\ub=\ps (u)$, now associated with this static observer, we will repeat the same steps that we did before for the definition of the vacuum state. (Here on we will replace the generic subscript `ob' by a subscript denoting the particular type of observer; in this case `so' stands for \emph{static observer.}) The world line of an observer keeping its radial position can be described as $r(\tau)=\rst=\cte$ and
\begin{equation}
t=\left(1-\frac{2m}{\rst}\right)^{-1/2} (\tau-\tau_0),
\label{t.of.r.static}
\end{equation}
so that  
\begin{equation}
\ub = \left(1-\frac{2m}{\rst}\right)^{-1/2} (\tau-\tau_0) - \rs (\rst).
\label{ub.of.u.static.t0}
\end{equation}
As before, an observer following this world line can naturally use its proper time $\tau$ to label the different rays he encounters while standing in its radial position, namely, he can make the assignment of the label $u = \ps^{-1} (\ub)=\tau$ given by the inverse of this equation to the ray $\ub=\cte$ that hits the observer at proper time $\tau$. Upon synchronization,  we obtain the relation between both labels
\begin{equation}
\ub = \ps(u) := \left(1-\frac{2m}{\rst}\right)^{-1/2} u.
\label{ub.of.u.static}
\end{equation}
Note that here one cannot demand $u \to \ub$ in  the past infinity, as we did in \sref{vacuum.choice}. The synchronization in this case reduces to eliminate  an irrelevant additive constant.

At this point, we can already construct the relation $U=U(u)$ we were after from 
the equation $\ub = \pv (U) = \ps (u)$. The complexity of~\eref{ub.of.U} does not 
allow us to write this relation explicitly, but we can write its inverse. 
Using~\eref{ub.of.U} and~\eref{ub.of.u.static}, we obtain
\begin{eqnarray}
\fl u = \left(1-\frac{2m}{\rst}\right)^{1/2} \left\{ \UH + \frac{4m}{3} - 4m \left[ \left( {3 \over 4m}(\UH-U) +1 \right)^{1/3}
\right. \right.
\nonumber \\ 
 + \left. \left.\ahalf \left( {3 \over 4m}(\UH-U) +1 \right)^{2/3} + \athird \left( {3 \over 4m}(\UH-U) +1 \right)
\right. \right.
\nonumber \\ 
+ \left. \left.\ln \left( \left({3 \over 4m}(\UH-U) +1\right)^{1/3}  - 1 \right) \right] \right\}.
\label{u.of.U.static}
\end{eqnarray}
One can study the behaviour of this function at the past and future infinities, $u \to -\infty$ and $u \to +\infty$, respectively. In the asymptotic past, the linear term on the right-hand side is the most important, and this relation becomes
\begin{equation}
U \approx \left(1-\frac{2m}{\rst}\right)^{-1/2} u, \qquad {\rm when} \quad u \to -\infty \qquad (U \to -\infty).
\label{u.of.U.static.past}
\end{equation}
By looking at the definition of $\kappa(u)$ in~\eref{kappa.def}, one can easily see that in this regime $\kappa=0$, so that there is not radiation at all.

On the other hand, at future infinity the behaviour of $U(u)$ is dominated by the logarithmic term in~\eref{u.of.U.static}, yielding an approximate expression
\begin{eqnarray}
\fl U \approx \UH -\frac{4m}{3} \left\{ \left[ \exp \left(\fr{4m} \left(\UH - 6m \right) \right)
\exp\left( -\left(1-\frac{2m}{\rst}\right)^{-1/2} \frac{u}{4m} \right) +1 \right]^3 - 1 \right\} 
\nonumber\\
\approx \UH - 4m \exp \left[\fr{4m} \left(\UH - 6m \right) \right] \exp\left[ -\left(1-\frac{2m}{\rst}\right)^{-1/2} \frac{u}{4m} \right],
\nonumber\\
{\rm when} \quad u \to \infty \qquad (U \to \UH).
\label{u.of.U.static.future}
\end{eqnarray}
From this expression, it is easy to recognize the well known asymptotic form~\cite{hawking2}
\begin{equation}
U=\UH-A\exp(-\kappa_{\rm a} u),
\label{asymptotic.formula}
\end{equation}
(where $\UH, A, \kappa_{\rm a}$ are constants), which is associated with a Planckian spectrum of particles with temperature $k_{\rm B} T=\kappa_{\rm a}/ (2\pi)$. Thus, when the static observer waits long enough, it perceives a Hawking flux with temperature 
\begin{equation}
k_{\rm B} T = \left(1-\frac{2m}{\rst}\right)^{-1/2}{1 \over 8\pi m}.
\label{T.static.observer}
\end{equation}
This is precisely the asymptotic Hawking temperature $k_{\rm B} T_{\rm H} = \kappa_{\rm H} / (2 \pi) = 1/(8\pi m)$, multiplied by a gravitational blue-shift factor which runs from unity for observers close to infinity, to arbitrarily large values for observers standing at positions closer and closer to the event horizon.

%------------------------------------
\subsubsection{Numerical \texorpdfstring{$\kappa(u)$}{kappa(u)}}
%------------------------------------

\begin{figure}[ht]
	\centering
    \includegraphics{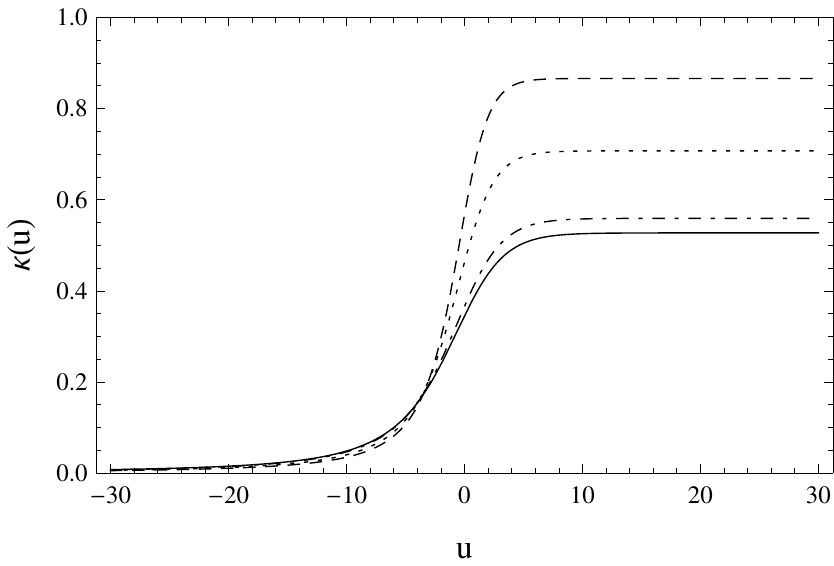}
  \caption{Temperature as a function of $u$ for different static observers at $\rst = $ ($3 m$, $4 m$, $10 m$, $20 m$) [curves depicted respectively as (\dashed, \dotted, \chain, \full)]. We use $2m=1$ units.}
  \label{kappa.static.graph}
\end{figure}

Let us describe now the particle perception of one of these observers throughout its 
entire life outside the horizon. From the definition of $\kappa(u)$ 
in~\eref{kappa.def}, it is easy to see that we can also write
\begin{equation}
\kappa (u) = \evat{\deriv[2]{u}{U} \left( \deriv{u}{U} \right)^{-2}}{U(u)}.
\label{alt.kappa}
\end{equation}
Then, evaluating the derivatives we can obtain a close form for $\kappa(u)$
\begin{equation}
\kappa (u) = \fr{4m} \left(1-\frac{2m}{\rst}\right)^{-1/2} \left\{ {3 \over 4m}[\UH-U(u)] +1 \right\}^{-4/3}.
\label{kappa.static}
\end{equation}
Solving numerically the relation $U(u)$ from~\eref{u.of.U.static}, we can plot $\kappa (u)$ for different static observers (\fref{kappa.static.graph}).

In these graphs, it is clearly seen how the vacuum state is such that in the past it does not contain particles, but at some instant of time it starts to heat up reaching asymptotically a final temperature. As we mentioned at the beginning, the vacuum state that we have chosen in a static background mimics what would happen in the formation of a black hole through gravitational collapse. Only once the black hole is very close to formation one enters in the asymptotic regime~\eref{u.of.U.static.future}, and Hawking radiation is switched on. We can estimate when this happen, but for that let us first describe the behaviour of the adiabatic control function in~\eref{adiabatic.condition}.

%------------------------------------------------------
\subsubsection{Validity of the adiabatic approximation}
%------------------------------------------------------

\begin{figure}[ht]
	\centering
		\includegraphics{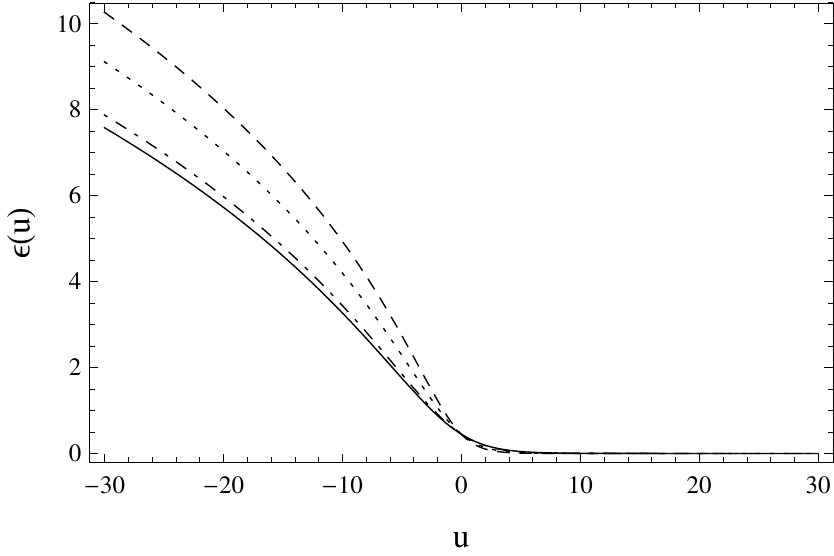}
  \caption{Adiabatic control function $\epsilon$ as a function of $u$ for different static observers at $\rst = $ ($3 m$, $4 m$, $10 m$, $20 m$) [curves depicted respectively as (\dashed, \dotted, \chain, \full)]. We use $2m=1$ units.}
  \label{epsilon.static.graph}
\end{figure}

From the definition of $\epsilon$ in~\eref{adiabatic.condition}, we have
\begin{equation}
\epsilon (u) = \evat{ \left| \deriv{\kappa}{U} \right| \left(\kappa^2 \deriv{u}{U} \right)^{-1}}{U(u)} = 4 \left\{\left[ {3 \over 4m}(\UH-U(u)) +1 \right]^{1/3} - 1 \right\}.
\label{epsilon.static}
\end{equation}
  Thus, $\epsilon$ decays from infinity at the asymptotic past, to zero when the 
observer that fixes the vacuum reaches the event horizon. This late time behaviour 
is consistent with the fact that then the quantum field is finally switched on, so that 
the radiation that a static observer sees is perfectly thermal. On the other hand, in 
the asymptotic past, $\epsilon$ becomes unbounded, but this does not mean a 
failure of the adiabatic condition, because $\kappa \to 0$ there. Again, we can  
numerically plot $\epsilon (u)$ for different static positions 
(\fref{epsilon.static.graph}). We can see that its value crushes to nearly zero when
 the temperature enters its asymptotic regime.

Now we can introduce two different referential times. On the one hand, we can define an \emph{ignition time} of the black hole as the moment at which $\ddot \kappa=0$. If one calculates this ignition time in terms of $U$, it does not depend on the value of $\rst$. This is perfectly reasonable, as the label $U$ corresponds to the single free-fall trajectory used to select the vacuum state. It is not difficult to see that this ignition time corresponds to $U_{\rm ig}= U_H-(676/1029)m$, or equivalently to the time when the free-fall trajectory reaches $r_{\rm ig} \simeq 2 m + 0.61 m$. On the other hand, we can define a \emph{thermality time} as the moment from which the outgoing radiation can be assumed to have a Planckian shape.
This thermality time can be calculated in terms of $U$ through the condition $\epsilon(U_{\rm th})=\epsilon_{\rm f}$, meaning that the adiabatic control function has reached a fiduciary value $\epsilon_{\rm f}$, that here we will take equal to $0.01$. This specific condition happens at $U_{\rm th} \simeq \UH - 0.01 m$, which corresponds to the time when the free-fall trajectory reaches $r_{\rm th} \simeq 2 m + 0.01 m$. Let us give you some figures. If a neutron star with 1.5 times the mass of the Sun (Schwarzschild radius equal to $4.43~\mathrm{km}$) and a radius of $12~\mathrm{km}$ started to undergo a free-fall collapse at Schwarzschild time $t_{\rm ini}$, it would reach $r_{\rm ig} \simeq 2m + 1.357~\mathrm{km}$ at a time $t_{\rm ig} \simeq t_{\rm ini} + 0.13 \sci{-3}{s}$, and $r_{\rm th} \simeq 2m + 0.022~\mathrm{km}$ at a time $t_{\rm th} \simeq t_{\rm ini} + 0.20 \sci{-3}{s}$. In this way, one can directly see that the detection of Hawking quanta at infinity does not involve any long delay associated with the freezing of the collapsing structure as it approaches the horizon formation.

Finally, let us comment that for different observers the previous two times are seen to happen at different instants of their proper times (their respective $u$ labels). The closer to the horizon, the later the observer will see Hawking radiation to switch on and to become Planckian.

%------------------------------------------------------------
\subsection{Observers freely falling from infinity}
\label{freely.falling.infinity}
%------------------------------------------------------------

In this subsection, we will analyse particle creation as seen by observers which are 
freely falling from infinity. To find the relevant re-labelling $\ub = \pf (u)$ 
(where `ffio' stands for \emph{falling-from-infinity observer}), we have to use the 
form of the radial time-like geodesic that starts at spatial infinity with zero velocity. 
However, recall that the calculation done to fix the vacuum state of the field in 
\sref{vacuum.choice} was precisely using these time-like trajectories. In that 
calculation, we had two constants of integration to play with, namely $t_0$ and 
$\tau_0$. We will now consider trajectories differing from the reference trajectory in 
\sref{vacuum.choice} by a temporal delay $\dt$. Instead of using 
$t_0=\tau_0=\UH+4m/3$, we will use new values 
$t_0=\tau_0=\UH+4m/3+\dt=\uH + 4m/3$, where $\uH := \UH + \dt$. We then 
define the new relation $\ub = \pf (u)$  by [see~\eref{ub.of.U}]
\begin{equation}
 \ub = \pf (u) := \pv(u-\dt)+\dt,
\label{ub.of.u.falling.impl}
\end{equation}
which explicitly reads
\begin{eqnarray}
\fl \ub = \pf (u) = \uH + \frac{4m}{3} - 4m \left\{ \left[ {3 \over 4m}(\uH-u) +1 \right]^{1/3} + \ahalf \left[ {3 \over 4m}(\uH-u) +1 \right]^{2/3}
\right. 
\nonumber \\ 
\left. 
+ \athird \left[ {3 \over 4m}(\uH-u) +1 \right] + \ln \left[ \left({3 \over 4m}(\uH-u) +1\right)^{1/3}  - 1 \right] \right\}.
\label{ub.of.u.falling}
\end{eqnarray}
Now, as we did for the static observer in \sref{static.observer}, we compare both labellings, $U$ in~\eref{ub.of.U} and $u$ in~\eref{ub.of.u.falling}, and obtain an implicit relation $U(u)$ from the equation
\begin{equation}
\pv (U) = \pf (u),
\label{U.of.u.falling.impl}
\end{equation}
which again explicitly reads
\begin{eqnarray}
\fl - 4m \left\{ \left[ {3 \over 4m}(\UH-U) +1 \right]^{1/3} + \ahalf \left[ {3 \over 4m}(\UH-U) +1 \right]^{2/3}
\right. 
\nonumber \\ 
\left. 
+ \athird \left[ {3 \over 4m}(\UH-U) +1 \right] + \ln \left[ \left({3 \over 4m}(\UH-U) +1\right)^{1/3}  - 1 \right] \right\}
\nonumber \\ 
= \dt - 4m \left\{ \left[ {3 \over 4m}(\uH-u) +1 \right]^{1/3} + \ahalf \left[ {3 \over 4m}(\uH-u) +1 \right]^{2/3}
\right. 
\nonumber \\ 
\left. 
+ \athird \left[ {3 \over 4m}(\uH-u) +1 \right] + \ln \left[ \left({3 \over 4m}(\uH-u) +1\right)^{1/3}  - 1 \right] \right\}.
\label{U.of.u.falling}
\end{eqnarray}

Solutions to this non-algebraic equation must be found numerically. Nevertheless, we can examine the past infinity and event horizon limits analytically. 
In the past infinity, we obtain $U \approx u$, owing to the synchronization we imposed. Near the event horizon, the logarithmic functions provide the leading contribution and therefore,
\begin{equation}
\fl U_{\rm near-hor} (u) \approx \UH
-\frac{4m}{3} \left\{\left[ \e^{-\dt/(4m)}\left( \left( {3 \over 4m}(\uH-u) +1 \right)^{1/3} - 1 \right) + 1 \right]^3 - 1 \right\}.
\label{U.of.u.falling.EH}
\end{equation}

The function $\kappa(u)$ can be readily obtained from~\eref{ub.of.u.falling.impl} 
and~\eref{U.of.u.falling.impl}. Indeed, taking the derivative 
of~\eref{U.of.u.falling.impl}
with respect to $u$, we obtain
\begin{equation}
\deriv{U}{u}  = \deriv{\pv(u-\dt)}{u} \left[\derivat{\pv(U)}{U}{U(u)}\right]^{-1},
\label{deriv.eq.in.f.falling}
\end{equation}
and differentiating once more
\begin{equation}
\fl \deriv[2]{U}{u} = \left[ \deriv[2]{\pv(u-\dt)}{u}  - 
\derivat[2]{\pv(U)}{U}{U(u)} \left( \deriv {U}{u}  \right)^2 \right] \left(\derivat{\pv(U)}{U}{U(u)}\right)^{-1},
\label{second.deriv.eq.in.f.falling}
\end{equation}
so that $\kappa(u)$ can be expressed as
\begin{equation}
\fl \kappa (u) = \left[ \derivat[2]{\pv(U)}{U}{U(u)} \left( \deriv{U}{u} \right)^2 - \deriv[2]{\pv(u-\dt)}{u} \right] \left(\deriv{\pv(u-\dt)}{u}\right)^{-1}.
\label{kappa.falling.derivs}
\end{equation}
Therefore, the final result is
\begin{eqnarray}
\fl \kappa (u) = \fr{4m} \left[ {3 \over 4m}(\uH-u) +1 \right]^{-1} \left\{ \left[{3 \over 4m}(\uH-u) +1 \right]^{1/3} - 1 \right\}^{-1} 
\nonumber\\
\times \left\{ \left[\frac{(3 / 4m)(\uH-u) + 1}{(3 / 4m)(\UH-U(u)) + 1}\right]^{4/3} - 1 \right\}.
\label{kappa.falling}
\end{eqnarray}

Note that this exact expression is explicit if we considered it as a function of both $u$ and $U$ (although {\em this does not mean that it is a function of two variables}). The dependence on $\dt$ is hidden in the implicit relation between both variables. We can use the limiting expression of $U(u)$ at the event horizon~\eref{U.of.u.falling.EH} (which is exact there) to find what temperature would the freely-falling observer characterized by the delay $\dt$ see when he crosses the horizon.
This temperature  happens to be
\begin{equation}
\kappa_{\rm HC}(\dt) = \fr{m} \left( 1- \rme^{-\dt/(4m)} \right).
\label{kappa.HC.falling}
\end{equation}
Therefore, the perceived radiation will run from an initial zero temperature to nearly $1/(2\pi m)$, which corresponds to four times the standard Hawking temperature, provided that the waiting time $\dt$ is several times bigger than $4m$. This is an interesting and, at first, puzzling result. How can one interpret it? In~\sref{Sec:interpretation}, we will offer and explanation of this result, but in order to do that we still need to introduce several additional material.

%------------------------------------------------------------
\subsubsection{Numerical \texorpdfstring{$\kappa(u)$}{kappa(u)}}
%------------------------------------------------------------

\begin{figure}[ht]
	\centering
		\includegraphics{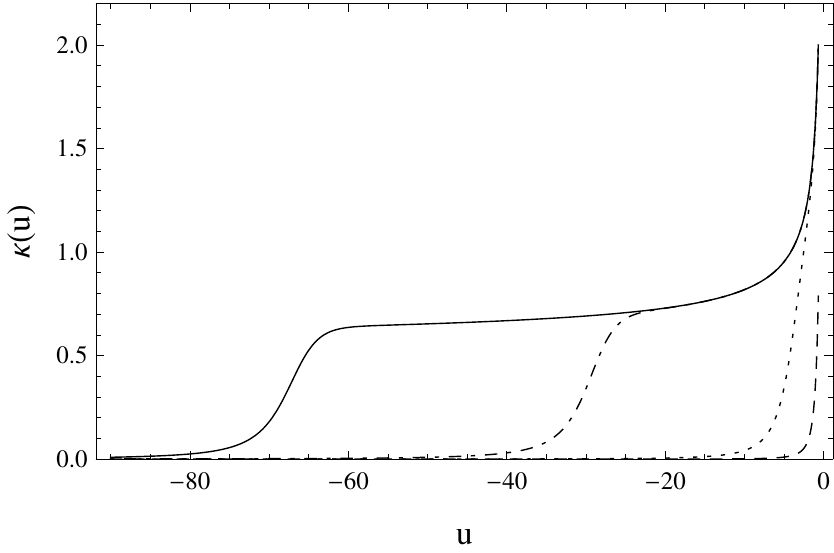}
  \caption{Temperature as a function of $u$ for different freely-falling observers from infinity with delays $\dt = $ ($2m$, $20m$, $100m$, $200m$) [curves depicted respectively as (\dashed, \dotted, \chain, \full)]. We use $2m=1$ units and $\uH = 0$.}
  \label{kappa.falling.graph}
\end{figure}

We can numerically plot the entire relation $\kappa(u)$. Results are shown in \fref{kappa.falling.graph}. For small values of $\dt$ (less than approximately $20 m$) the temperature perceived by the freely-falling observer is nearly zero during most of its trajectory towards the horizon. Only when approaching the horizon, the temperature increases to reach a peak value given by~\eref{kappa.HC.falling}. For large enough values of $\dt$, before this final increase in the temperature, there appears an intermediate plateau in which the temperature is nearly constant. It is easy to understand this behaviour: if the freely-falling observer is still far from the horizon when Hawking radiation is switched on, at some moment it will start detecting this radiation, which will stay nearly constant till the observer becomes close to the horizon, where additional physics will come into play (see \sref{Sec:interpretation}). The effective temperature value along the plateau is somewhat higher than the Hawking radiation temperature, and slightly increasing. As we will see, this is mostly due to a Doppler blue-shift associated with the radial velocity of the falling observer (see also \sref{Sec:interpretation}).

\begin{figure}[ht]
	\centering
		\includegraphics{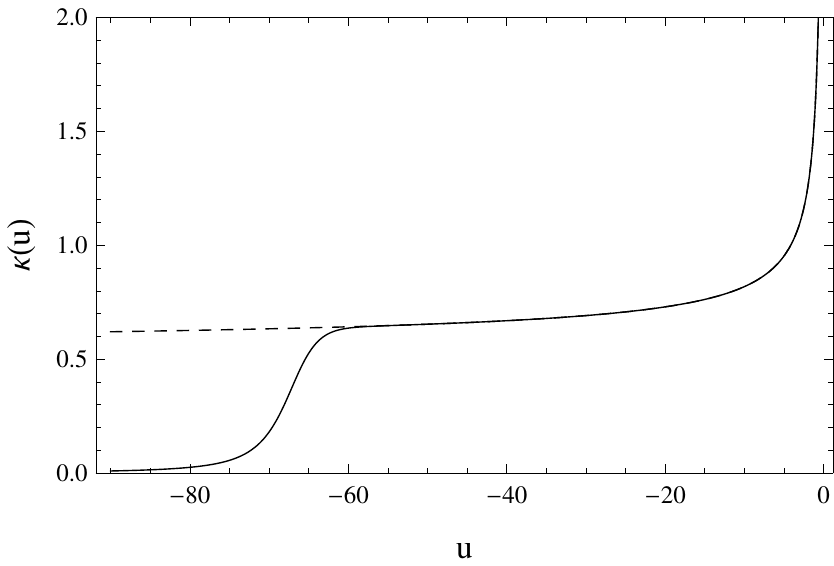}
  \caption{Temperature as a function of $u$ for a freely-falling observer from infinity with $\dt = 200 m$ using the exact expression (curve depicted \full) and using $U_{\rm near-hor} (u)$ approximation (curve depicted \dashed). We use $2m=1$ units and $\uH = 0$.}
  \label{eh.falling.graph}
\end{figure}

It is interesting to look at how good is the functional approximation~\eref{U.of.u.falling.EH} when plugged into~\eref{kappa.falling}, as compared with the exact result. This comparison is shown in \fref{eh.falling.graph} for a long delay $\dt$. By construction, this approximation works perfectly well in the surroundings of the horizon crossing. However, it is remarkable that it also fits very well the plateau region.

\begin{figure}[ht]
	\centering
		\includegraphics{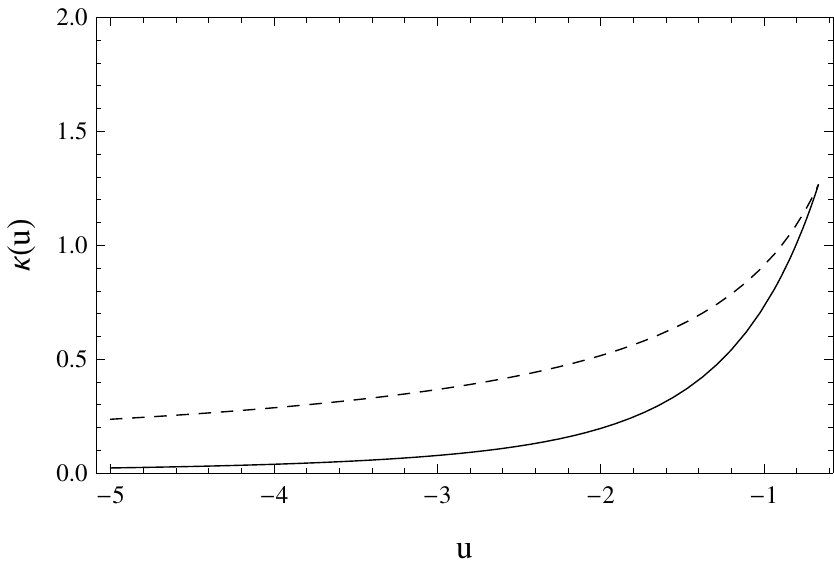}
  \caption{Temperature as a function of $u$ for a freely-falling observer from infinity with $\dt = 4 m$ using the exact expression (curve depicted \full) and using $U_{\rm near-hor} (u)$ approximation (curve depicted \dashed). We use $2m=1$ units and $\uH = 0$.}
  \label{low.falling.graph}
\end{figure}

On the other hand, this approximation is not useful when $\dt$ is not large enough (see \fref{low.falling.graph}). The asymptotic approximation~\eref{U.of.u.falling.EH} correctly reproduces the horizon-crossing value as should be by construction, but at early times it deviates from the exact curve rather quickly.

%---------------------------------------------------------------
\subsubsection{Adiabatic approximation validity}
%---------------------------------------------------------------

Having found an expression for $\kappa$ in terms of $u$ and $U(u)$, we can use it 
to obtain an expression for the adiabatic control function $\epsilon(u)$ as defined 
in~\eref{adiabatic.condition}. The 
resulting expression, also explicit when considered as a function of both $u$ and $U$, is rather involved, and we will not include it here. But, as we did with $\kappa$, we can also replace the relation $U_{\rm near-hor}(u)$~\eref{U.of.u.falling.EH} 
valid in the near-horizon limit to find an explicit expression for $\epsilon(u)$ valid there; finally, we can take the
 $u \to \uH$ limit and find its value at horizon-crossing. The result is
\begin{equation}
\epsilon_{\rm HC}(\dt) = \frac{3}{8} + \frac{7}{4 \left[ \e^{\dt/(4m)} - 1 \right]}.
\label{epsilon.falling.EH}
\end{equation}
This expression runs from $3/8$ when $\dt$ is several times larger than $4m$, 
growing to infinity when $\dt$ goes to zero. This last value is comprehensible from 
the definition of $\epsilon$, as it is the temperature itself that goes to zero when the
 freely-falling observer approaches the referential one. The adiabatic condition 
$\epsilon \ll 1$ is never satisfied at horizon crossing. However, it is remarkable that, 
for $\dt$ sufficiently large, it is always still smaller than unity, so that the perceived 
particle spectrum should not be very different from a Planckian 
spectrum.

\begin{figure}[ht]
	\centering
		\includegraphics{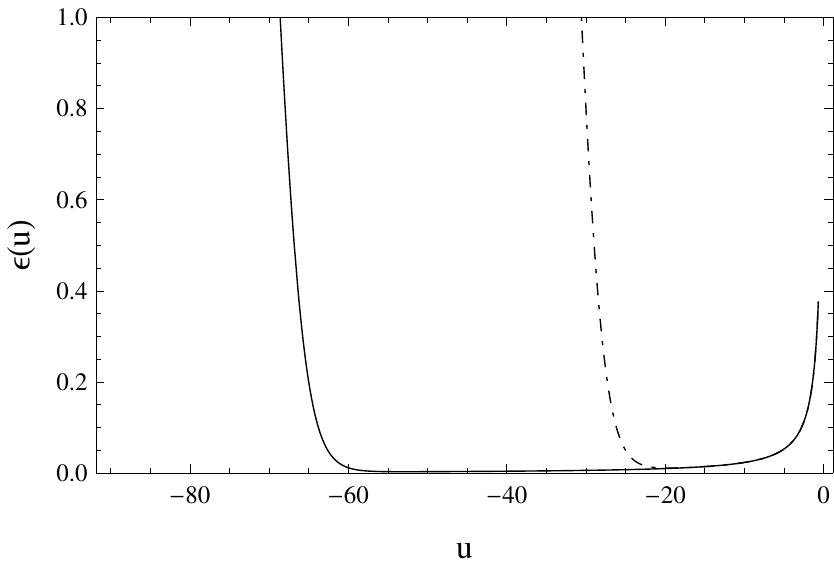}
  \caption{Adiabatic control function $\epsilon$ as a function of $u$ for different freely-falling observers from infinity with delays $\dt = $ ($100m$, $200m$) [curves depicted respectively as (\chain, \full)]. We use $2m=1$ units and $\uH = 0$.}
  \label{epsilon.falling.graph}
\end{figure}

Nonetheless, this is just the horizon crossing value. We can see using the numerically evaluated exact result (\fref{epsilon.falling.graph}) that, for long waiting times, there is part of the trajectory for which the adiabatic condition is with no doubt valid, as there $\epsilon$ crushes to nearly zero. This of course coincides with the plateau region (in the cases in which it appears), in which $\kappa$ takes a nearly constant value. Once the final increase in $\kappa$ shows up, the adiabatic condition starts to be violated, although in a rather mild manner.

%---------------------------------------------------------------
\subsection{Observers freely falling from a finite radius}
\label{freely.falling.radius}
%---------------------------------------------------------------

The last situation we shall consider in this paper is the particle perception of a 
freely-falling observer left to fall from a radius $r_0$, with zero initial velocity at 
some waiting time $\dtw$ (in Schwarzschild time) after the reference trajectory 
used to define the vacuum state had crossed this radial position. Again, we 
start by calculating this new trajectory. For that, we have to 
integrate~\eref{r.tau.first.integral}, but now keeping $r_0$ a finite constant. This 
yields
\begin{equation}
{\tilde\tau}_0 -\tau= - r_0 \sqrt{r_0 \over 2m } \left\{ \left( \frac{r_0}{r} - 1 \right)^{1/2} \frac{r}{r_0} + \arctan \left[\left( \frac{r_0}{r} - 1 \right)^{1/2} \right] \right\},
\label{tau.of.r.radius}
\end{equation}
where ${\tilde\tau}_0$ represents the time at which the observer starts to fall (note that, in all expressions in this subsection, $2m < r \leq r_0$). The time of horizon-crossing for this trajectory is
\begin{equation}
{\tilde u}_{\rm H}:= {\tilde\tau}_0 + 2m \sqrt{r_0 \over 2m } \left\{ \left( \frac{r_0}{2m} - 1 \right)^{1/2} + \frac{r_0}{2m} \arctan \left[\left( \frac{r_0}{2m} - 1 \right)^{1/2} \right] \right\}.
\label{u_H.radius}
\end{equation}
By means of~\eref{r.tau.first.integral} and~\eref{t.tau.first.integral}, we obtain
\begin{equation}
\deriv{r}{t} = -\sqrt{\frac{2m}{r}}\left(1- {2m \over r}\right) \left(1  -\frac{r}{\rc} \right)^{1/2}
\left(1  -\frac{2m}{\rc} \right)^{-1/2},
\label{dr.of.t.radius}
\end{equation}
which can be readily integrated to yield
\begin{eqnarray}
\fl t (r) = {\tilde t}_0 + r_0 \left( \frac{r_0}{2m} - 1 \right)^{1/2} \left\{ \left( \frac{r_0}{r} - 1 \right)^{1/2} \frac{r}{r_0} + \left( 1 + \frac{4m}{r_0} \right) \arctan \left[\left( \frac{r_0}{r} - 1 \right)^{1/2} \right] \right\}
\nonumber\\
+ \ln \left\{  \left[ \frac{r}{2m} + 2 \left(1- \frac{2m}{r_0}\right)^{1/2} \left(1- \frac{r}{r_0}\right)^{1/2} \sqrt{\frac{r}{2m}} - \frac{2 r}{r_0} + 1 \right]
\right.
\nonumber\\
\left.
\times \left( \frac{r}{2m} - 1\right)^{-1} \right\},
\label{t.of.r.radius}
\end{eqnarray}
where ${\tilde t}_0$ is again an integration constant, in this case representing the time at which the observer starts to fall from $r_0$. If we call $t_{\rm vs} (r)$ the expression in~\eref{t.of.r}, we can define ${\tilde t}_0$ as a function of $r_0$, $t_0$ and the delay time $\dtw$
\begin{eqnarray}
\fl {\tilde t}_0 = t_{\rm vs} (r_0) + \dtw
\nonumber \\
= t_0 -4m \left[ \sqrt{\frac{r_0}{2m}} + \athird \left( \frac{r_0}{2m} \right)^{3/2} + \ahalf \ln \left( \frac{\sqrt{r_0/(2m)}-1}{\sqrt{r_0/(2m)}+1} \right) \right] + \dtw.
\label{delay.fixing.radius}
\end{eqnarray}
Substituting in~\eref{t.of.r.radius} the function $r(\tau)$ which is implicitly defined in~\eref{tau.of.r.radius}, we can find the function $t(\tau)$. 

Let us now construct the spacetime trajectory of an observer that 
remains 
at a fixed position $r=r_0$ until $\tau=\tilde\tau_0$, at which time he starts to fall 
freely towards the black hole. This trajectory $(t(\tau), r(\tau))$ is made out of 
two pieces jointed together at $\tau={\tilde\tau}_0$: the first piece, for $\tau < 
{\tilde\tau}_0$, is such that the radius is fixed $r=r_0$
and the Schwarzschild time satisfies
\begin{equation}
t(\tau)= {\tilde t}_0 + \left(1-{2m \over r_0}\right)^{-1/2}(\tau - {\tilde\tau}_0);
\end{equation}
the second piece, for $\tau \geq {\tilde\tau}_0$, is constituted by the previously found implicit expressions $r(\tau;r_0,{\tilde\tau}_0)$ in \eref{tau.of.r.radius} and $t(r(\tau;r_0,{\tilde\tau}_0);r_0, t_0,\dtw)=t(\tau;r_0,{\tilde\tau}_0,t_0,\dtw)$ as defined in~\eref{t.of.r.radius}; the $t_0$ and $\dtw$ dependence of this last expression comes from the dependence of ${\tilde t}_0$ on $t_0$ and $\dtw$ in~\eref{delay.fixing.radius}.

As explained in previous sections, substituting $\tau$ by $u$, using an appropriate synchronization between $t_0$ and ${\tilde\tau}_0$, and substituting ${\tilde\tau}_0$ by $\tilde{u}_{\rm H}$ as defined in~\eref{u_H.radius}, we can construct the function
\begin{equation}
\ub=p_{\rm ffro}(r):=t(r;r_0,\tilde{u}_{\rm H},\dtw)-r^*(r),
\end{equation}
where `ffro' stands for \emph{freely-falling-from-a-radius observer}.
Finally, the $U(u)$ relation we were looking for can be found from the implicit relation
\begin{equation}
\bar u=\pv (U) = \pr (r(u)).
\label{eq.in.f.radius}
\end{equation}

Our next step is to find the form of $\kappa(u)$ for these observers.
Taking derivatives of \eref{eq.in.f.radius} with respect to $u$, we obtain
\begin{equation}
\deriv{U}{u} = \left( \deriv{\pv}{U} \right)^{-1} \deriv{\pr}{r} \deriv{r}{u},
\label{deriv.eq.in.f.radius}
\end{equation}
where $\rmd r / \rmd u$ is nothing but~\eref{r.tau.first.integral} with $u$ instead of $\tau$, and differentiating again,
\begin{equation}
\deriv[2]{U}{u} = \left[ \deriv[2]{\pr}{r} \left( \deriv{r}{u} \right)^2 + \deriv{\pr}{r} \deriv[2]{r}{u} - \deriv[2]{\pv}{U} \left( \deriv{U}{u} \right)^2 \right] \left( \deriv{\pv}{U} \right)^{-1}.
\label{second.deriv.eq.in.f.radius}
\end{equation}
In this way, $\kappa(u)$ can be expressed as
\begin{equation}
\kappa(u) = \left( \deriv{\pv}{U} \right)^{-1} \deriv[2]{\pv}{U} \deriv{U}{u} - \left( \deriv{\pr}{r} \right)^{-1} \deriv[2]{\pr}{r} \deriv{r}{u} - \left( \deriv{r}{u} \right)^{-1} \deriv[2]{r}{u}.
\label{kappa.radius.derivs}
\end{equation}
Substituting explicitly the corresponding expressions and simplifying, it becomes
\begin{eqnarray}
\fl \kappa(u) = \fr{4m} \left(1-\frac{2m}{r_0}\right)^{-1/2} \left[ {3 \over 4m}(\UH-U) +1 \right]^{-4/3}, \qquad {\rm for} \quad u< u_0,
\label{kappa.static.radius}
\\
\fl \kappa(u) = \fr{4m} \left\{\frac{r}{2m}+1 +\frac{r}{m} \left[ \left(1-\frac{2m}{r_0}\right)^{1/2} \sqrt{\frac{2m}{r}} \left(1-\frac{r}{r_0}\right)^{1/2}  - \frac{2m}{r_0} \right]\right\}
\nonumber\\
\times \left\{\left[ \left(1-\frac{2m}{r_0}\right)^{1/2} + \sqrt{\frac{2m}{r}}\left(1-\frac{r}{r_0}\right)^{1/2} \right] \left(\frac{r}{2m}-1\right)\right\}^{-1}
\nonumber\\
\times \left\{\left[\frac{3}{4m}(\UH-U)+1\right]^{-4/3} - \left(\frac{2m}{r}\right)^2 \right\},
\qquad {\rm for} \quad u \geq u_0,
\label{kappa.radius}
\end{eqnarray}
where $r$ and $U$ must be understood as functions of $u$  (again, these can only be obtained numerically). This function has a finite jump at $u=u_0~(= {\tilde\tau}_0)$, the moment at which the observer at $r_0$ is released from its previously fixed position.

At early times, $\kappa$ follows the same pattern described in \sref{static.observer}, in~\eref{kappa.static}. It increases from zero and tries to reach the asymptotic value associated with the observer at rest at position $r_0$. However, at some moment $u_0$ the observer is released to fall freely. In this moment, $\kappa$ undergoes a finite jump. The value of $\kappa$ just after the jump, which we will denote $\kappa_{\rm released}$, is found by evaluating \eref{kappa.radius} at $r=r_0$, which gives
\begin{eqnarray}
\fl \kappa_{\rm released}(r_0,\dtw) = \fr{4m} \left(1 - \frac{2m}{r_0} \right)^{-1/2}
\nonumber\\
\times \left\{ \left[ \frac{3}{4m} (\UH - U_0(r_0,\dtw)) + 1 \right]^{-4/3} - \left(\frac{2m}{r_0}\right)^{-2} \right\}.
\label{kappa.init.radius}
\end{eqnarray}
Here $U_0 := U(u_0)$ is the value of the label $U$ associated with $u_0$, and depends on the parameter $\dtw$. A particular case is $\dtw = 0$, which means that the observer starts to fall just when the observer that fixes the vacuum state \emph{passes by its side.} This implies of course $U_0 = U(r_0)$, as determined by the trajectory in \sref{vacuum.choice} [it can be obtained by reversing~\eref{r.of.tau}, noting that $U$ acts as $\tau$ now]. Doing this one can check that $\kappa_{\rm released}$ is zero, as one should expect. This is because in this case, at the starting point, between the \emph{observer that determines the vacuum state} and our observer there is no difference but in their velocity. If the former one sees nothing, the latter shall see ``nothing times a red-shift factor'', that is, \emph{nothing at all.}

On the other hand, if the observer  waits long enough so that the black hole is nearly completely switched on (Unruh state),
that is, $\dtw \to \infty$, which amounts to substituting $U$ by $\UH$ in equation~\eref{kappa.init.radius}, then $\kappa_{\rm released}$ takes the value
\begin{equation}
\kappa_{\rm Unruh} (r_0):=\evat{\kappa_{\rm released}(r_0)}{\dtw \to \infty} = \fr{4m} \left(1 +\frac{2m}{r_0}\right) \left(1 -\frac{2m}{r_0}\right)^{1/2}.
\label{kappa.Unruh.radius}
\end{equation}
The limit $r_0 \to 2m$ gives zero, and the limit $r_0 \to \infty$ gives the usual Hawking temperature $\kappa = 1/(4m)$, as happens in the Unruh state. This agrees with the fact that our state reproduces the Unruh state for late times.

\begin{figure}[ht]
	\centering
		\includegraphics{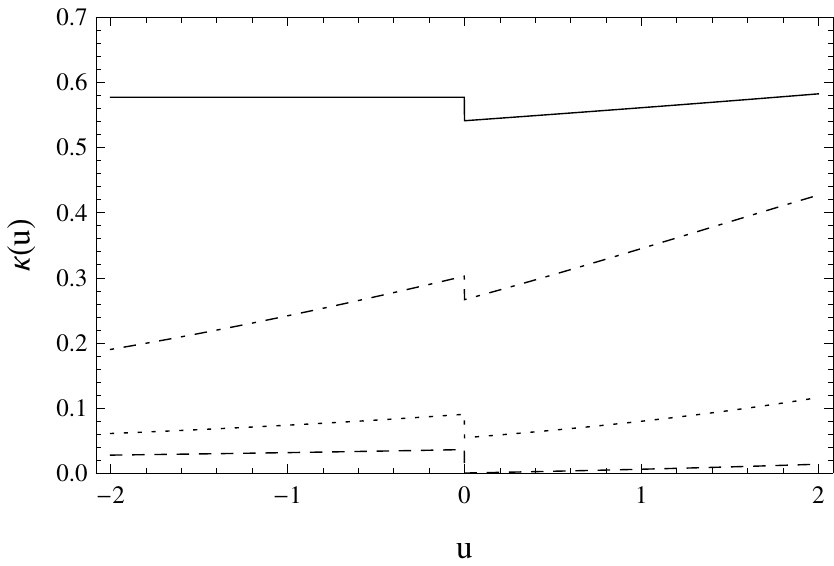}
  \caption{Temperature around the releasing point as a function of $u$ for $r_0 = 8m$ and delays $\dtw =$ ($0$, $12 m$, $24m$, $\infty$) [curves depicted respectively as (\dashed, \dotted, \chain, \full)]. We use $2m=1$ units and $u_0 = 0$.}
  \label{kappa.jump.radius.graph}
\end{figure}

Another interesting quantity we can calculate is the value of the jump in $\kappa$, $\Delta \kappa_{\rm release}$, at the instant of releasing.
We can directly subtract from~\eref{kappa.static.radius} the expression in~\eref{kappa.init.radius}, and find the \emph{jump} $\Delta \kappa_{\rm release}$,
\begin{equation}
\Delta \kappa_{\rm release} (r_0)= \fr{4m} \left(1-{2m \over r_0}\right)^{-1/2}\left({2m \over r_0}\right)^2=
\left(1-{2m \over r_0}\right)^{-1/2} {m \over r_0^2}.
\label{jump.kappa}
\end{equation}
It is remarkable that it does not depend on $U$ or, in other words, does not depend on the waiting time $\dtw$. This is reasonable, as by \emph{switching off the rockets} the observer is just loosing the radiation that comes from its own acceleration, and this acceleration only depends on the radius it was staying at. In fact, in the last expression we can directly spot the surface gravity associated with position $r_0$ [see~\eref{Newton}] multiplied by its corresponding gravitational blue-shift factor. We can conclude then that $\Delta \kappa_{\rm release}$ is nothing but Unruh radiation associated with an accelerating observer~\cite{unruh, birrell-davies}. In \fref{kappa.jump.radius.graph}, we plot the value of $\kappa$ around the release point for different waiting times. There, one can directly see that the jump in the temperature is always the same.

Now, let us study the radiation perception at horizon crossing. The procedure is analogue to the one followed in \sref{freely.falling.infinity}: in~\eref{eq.in.f.radius} we identify the diverging terms at the event horizon and, from the resulting equation, isolating $U$ we find an expression $U_{\rm near-hor} (r)$ valid near the event horizon. The expression is quite complex and not really useful by itself, but again, plugging it into~\eref{kappa.radius} we obtain an expression $\kappa(r)$ valid near the event horizon. Finally, performing the limit $r \to 2m$ we find the horizon crossing value $\kappa_{\rm HC}$, which in this case is 
\begin{eqnarray}
\fl \kappa_{\rm HC}(r_0,\dtw) = \fr{m} \left\{ \left(1-\frac{2m}{r_0}\right)^{1/2} - \left( \frac{\sqrt{r_0/(2m)} - 1}{\sqrt{r_0/(2m)} + 1} \right)^{1/2}
\right.
\nonumber\\
\times \left.\exp \left[ -\frac{5}{6} - \frac{r_0+\dtw}{4m} + \sqrt{\frac{r_0}{2m}} + \athird \left( \frac{r_0}{2m} \right)^{3/2}
\right. \right.
\nonumber\\
\left. \left. -\left(\frac{r_0}{4m} + 1 \right) \left( \frac{r_0}{2m} - 1 \right)^{1/2} \arctan \left( \left( \frac{r_0}{2m} - 1 \right)^{1/2} \right) \right]\right\}.
\label{kappa.HC.radius}
\end{eqnarray}
When the observer waits for a sufficiently long time before being released from its position (mathematically, $\dtw \to \infty$), it simplifies to
\begin{equation}
\evat{\kappa_{\rm HC}}{\dtw \to \infty} = \fr{m} \left(1-\frac{2m}{r_0}\right)^{1/2},
\label{kappa.HC.Unruh.radius}
\end{equation}
which reproduces the result $\kappa = 1 / m$ for an observer freely falling from infinity [see the $\dt \to \infty$ limit of~\eref{kappa.HC.falling}]. It is also interesting to analyse the limit of $\kappa_{\rm HC} (r_0, \dtw)$ when $r_0 \to \infty$. The result is 
\begin{equation}
\evat{\kappa_{\rm HC}}{r_0 \to \infty} = \fr{m},
\label{kappa.HC.infinity.radius}
\end{equation}
which, surprisingly at first sight, does not depend on $\dtw$. That is, we cannot reproduce the entire formula~\eref{kappa.HC.falling} by taking the $r_0 \to \infty$ limit. The reason is that the trajectory used to fix the vacuum state does never coincide with the trajectory of a released observer even if the release point $r_0$ is taking close to infinity. Although the velocity difference of these two trajectories at $r_0$ tends to zero when $r_0 \to \infty$, they arrive at positions close to the horizon at very different Schwarzschild times. The freely-falling-from-a-radius observer for far enough releasing radius arrives to the horizon always so late with respect to the reference trajectory, that it sees the black hole as virtually switched on (in its final Unruh state).

\begin{figure}[ht]
	\centering
		\includegraphics{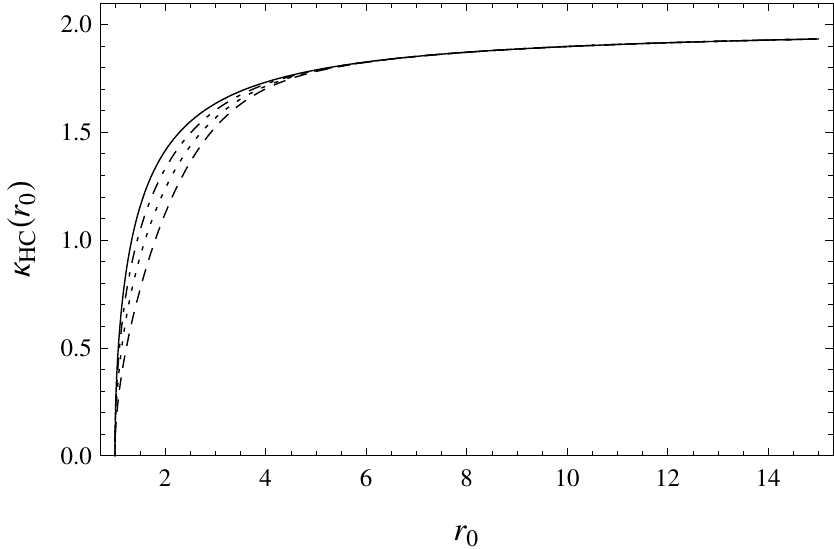}
  \caption{Temperature when crossing the horizon as a function of $r_0$ for delays $\dtw =$ ($0$, $2 m$, $5 m$, $\infty$) [curves depicted respectively as (\dashed, \dotted, \chain, \full)]. We use $2m=1$ units.}
  \label{kappa.HC.radius.graph}
\end{figure}

Plotting expression~\eref{kappa.HC.radius} (\fref{kappa.HC.radius.graph}), we can see that the parameter $\dtw$ plays almost no role in the temperature seen when crossing the horizon.

%-------------------------------------------------
\subsubsection{Numerical \texorpdfstring{$\kappa(u)$}{kappa(u)}}
\label{numerical.kappa.radius}
%-------------------------------------------------

\begin{figure}[ht]
	\centering
		\includegraphics{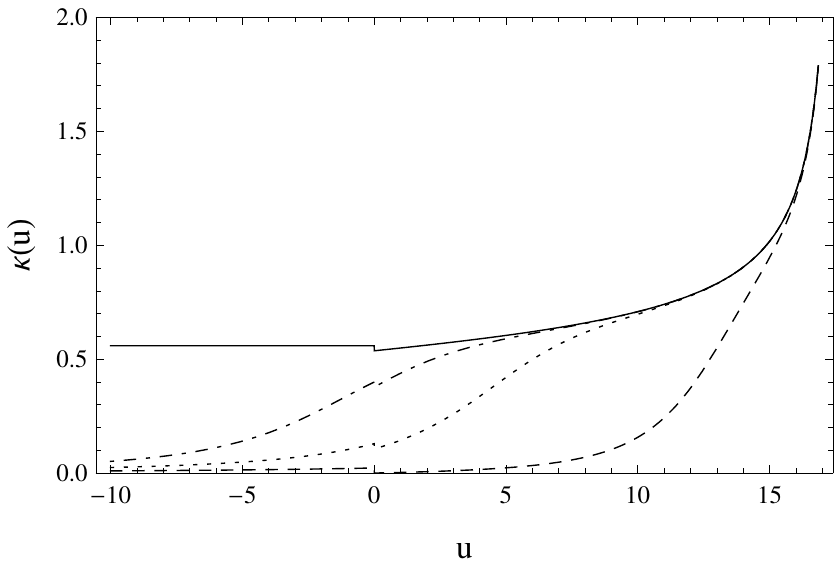}
  \caption{Temperature as a function of $u$ for $r_0 = 10 m$ and delays $\dtw =$ ($0$, $24 m$, $36 m$, $\infty$) [curves depicted respectively as (\dashed, \dotted, \chain, \full)]. We use $2m=1$ units and $u_0 = 0$.}
  \label{kappa.times.radius.graph}
\end{figure}

Let us now plot some examples of exact numerical resolution of 
$\kappa(u)$ for different trajectories. The aim is just to have a visual 
and qualitative understanding of the global evolution of the temperature, as well as 
to see how the calculated exact limits and behaviours are reproduced. First, we will 
plot the value of $\kappa(u)$ for observers starting to fall from the radius $r_0 = 
10m$ with different waiting times (\fref{kappa.times.radius.graph}). Graphs show 
how $\kappa(u)$ starts from very different values when the observer is released, 
but converges to virtually the same final value no matter the waiting time, as the 
starting radius is far enough.

\begin{figure}[ht]
	\centering
		\includegraphics{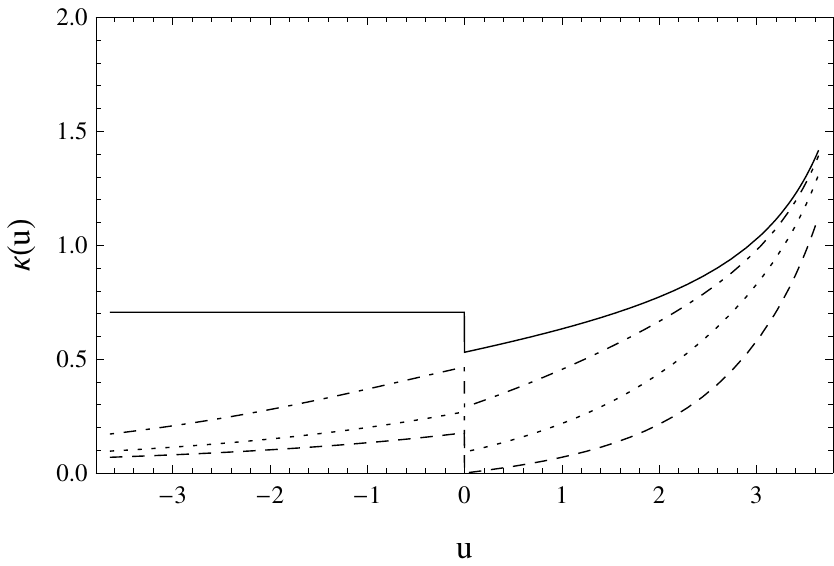}
  \caption{Temperature as a function of $u$ for $r_0 = 4 m$ and delays $\dtw =$ ($0$, $4 m$, $10 m$, $\infty$) [curves depicted respectively as (\dashed, \dotted, \chain, \full)]. We use $2m=1$ units and $u_0 = 0$.}
  \label{kappa.times2.radius.graph}
\end{figure}

If the radius $r_0$ is quite near the event horizon ($r_0 = 4m$ in \fref{kappa.times2.radius.graph}), there is a small but appreciable difference in the value of $\kappa(u)$ when crossing it, depending on the waiting time. We can also see that the jump in $\kappa(u)$ when releasing the observer is more important in this case, as acceleration to stay at the fixed radius is much bigger here.

\begin{figure}[ht]
	\centering
		\includegraphics{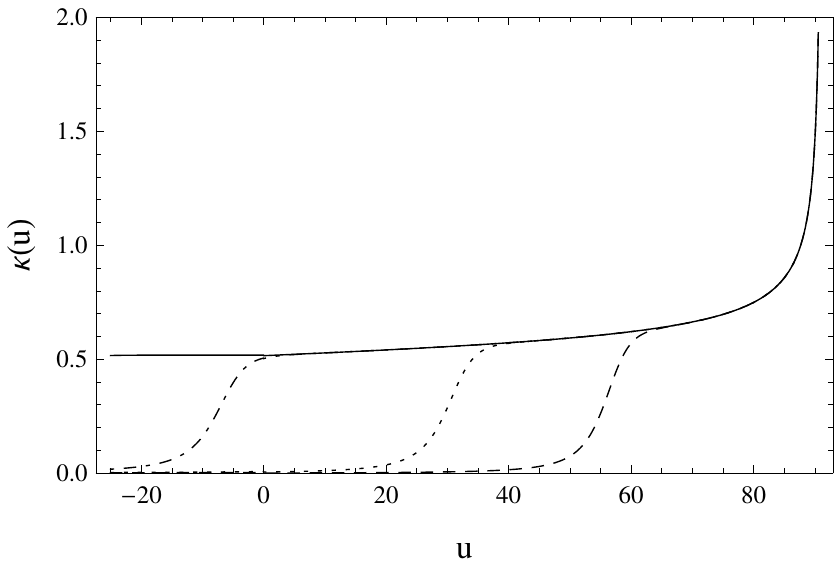}
  \caption{Temperature as a function of $u$ for $r_0 = 30 m$ and delays $\dtw =$ ($0$, $60 m$, $140 m$, $\infty$) [curves depicted respectively as (\dashed, \dotted, \chain, \full)]. We use $2m=1$ units and $u_0 = 0$.}
  \label{kappa.times3.radius.graph}
\end{figure}

On the other hand, if the radius is really far from the event horizon ($r_0 = 30m$ in \fref{kappa.times3.radius.graph}), the jump in $\kappa(u)$ is completely negligible. Different waiting times in this case just determine when you start perceiving the radiation, which can happen before or after the releasing point. The plateau found in \sref{freely.falling.infinity}, and of course the final peak, are again reproduced, as should happen for a large radius. But note that, as we already argued, it is not possible to reproduce the cases in \sref{freely.falling.infinity} for which $\dt$ is small. The limit $r_0 \to \infty$ here yields the same limit than $\dt \to \infty$ in \sref{freely.falling.infinity}, no matter the value of $\dtw$ here.

%-------------------------------------------------
\subsubsection{Validity of the adiabatic approximation}
%-------------------------------------------------

\begin{figure}[ht]
	\centering
		\includegraphics{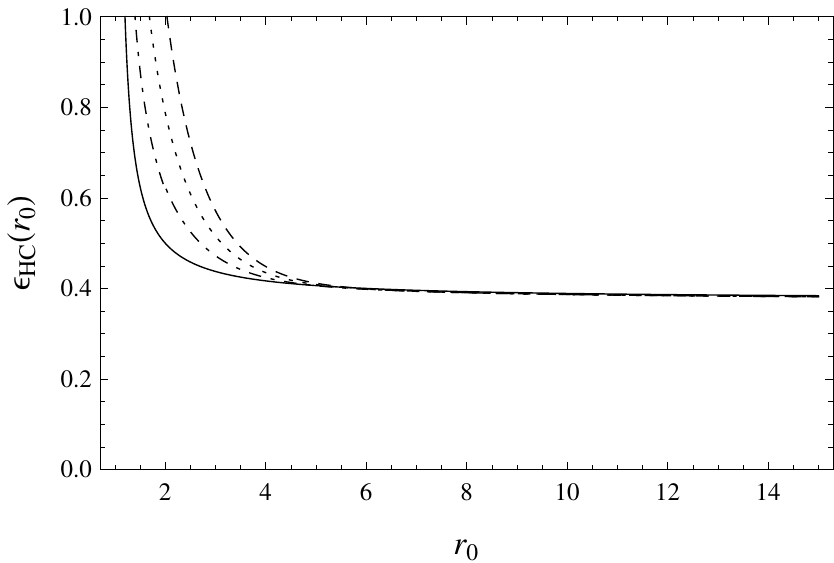}
  \caption{Adiabatic control function $\epsilon$ when crossing the horizon as a function of $r_0$ for delays $\dtw =$ ($0$, $2 m$, $5 m$, $\infty$) [curves depicted respectively as (\dashed, \dotted, \chain, \full)]. We use $2m=1$ units.}
  \label{epsilon.HC.radius.graph}
\end{figure}

Repeating what we have done for the observers studied so far, here we can find a formula for the adiabatic condition $\epsilon(u)$, in this case as an explicit function of both $U=U(u)$ and $r=r(u)$. Also, once more we can plug the expression $U_{\rm near-hor} (r)$ valid near the event horizon, and find a function $\epsilon (r)$ valid there. Finally, we may take the limit $r \to 2m$ and obtain the value $\epsilon_{\rm HC}$ at the horizon crossing. Unfortunately, in this case the approximation $U_{\rm near-hor} (r)$, and the exact expression $\epsilon (U,r)$ are extremely complex in their functional forms. Thus, it seems neither possible, nor interesting, to find an explicit expression of $\epsilon_{\rm HC}$ in terms of the parameters $r_0$ and $\dtw$. Nonetheless, it is still possible to obtain an expression for $\epsilon (r, r_0)$ (not only in the horizon crossing) when $\dtw \to \infty$,
\begin{eqnarray}
\fl \evat{\epsilon (r, r_0)}{\dtw \to \infty} = \left[ 3 - \frac{4r}{r_0} + \left(\frac{r}{2m}\right)^2 - 4 \left(1-\frac{2m}{r_0}\right)^{1/2}  \left(1-\frac{r}{r_0}\right)^{1/2} \sqrt{\frac{r}{2m}} \right]
\nonumber\\
\times \left[ \left(\frac{r}{2m}\right)^2 -1 \right]^{-2}.
\label{epsilon.radius}
\end{eqnarray}
In this expression, one can now take the limit $r \to 2m$ and obtain $\epsilon_{\rm HC}$ when $\dtw \to \infty$ in terms of $r_0$,
\begin{equation}
\evat{\epsilon_{\rm HC} (r_0)}{\dtw \to \infty} = \fr{8} \left[ 3 + \left( \frac{r_0}{2m} - 1 \right)^{-1} \right].
\label{epsilon.HC.radius}
\end{equation}
This result runs from $3/8$ when $r_0 \gg 2m$, to infinity when $r_0 \to 2m$. With the first limit $3/8$, we reproduce again the result obtained in \sref{freely.falling.infinity} for long delays [see the limit of~\eref{epsilon.falling.EH}]. The second divergent limit reflects the fact that observers released from positions near the horizon by no means perceive thermal radiation. In \fref{epsilon.HC.radius.graph}, we plot the function in~\eref{epsilon.HC.radius}, together with some numerically solved graphs of $\epsilon_{\rm HC}$ for zero and finite waiting times $\dtw$. Note that the limit $3/8$ when $r_0 \to \infty$ is common for all the graphs.

Also, we can find an expression for $\epsilon_{\rm released}$ at the beginning of the falling part of the path, again for $\dtw \to \infty$. We obtain
\begin{equation}
\evat{\epsilon_{\rm released} (r_0)}{\dtw \to \infty} = \left[ \left(\frac{r_0}{2m}\right)^2 - 1 \right]^{-1}.
\label{epsilon.init.radius}
\end{equation}
This yields zero for $r_0 \gg 2m$, and infinity when $r_0 \to 2m$. The zero value appears because, at large radius and for large waiting times, what the observer initially sees is a perfect Hawking radiation. The divergent value for $r_0$ near the horizon has the same explanation as in the case of $\epsilon_{\rm HC}$: for trajectories starting too close to the horizon, the adiabatic approximation is never valid. Neither at the beginning of the fall, nor at horizon crossing.

\begin{figure}[ht]
	\centering
		\includegraphics{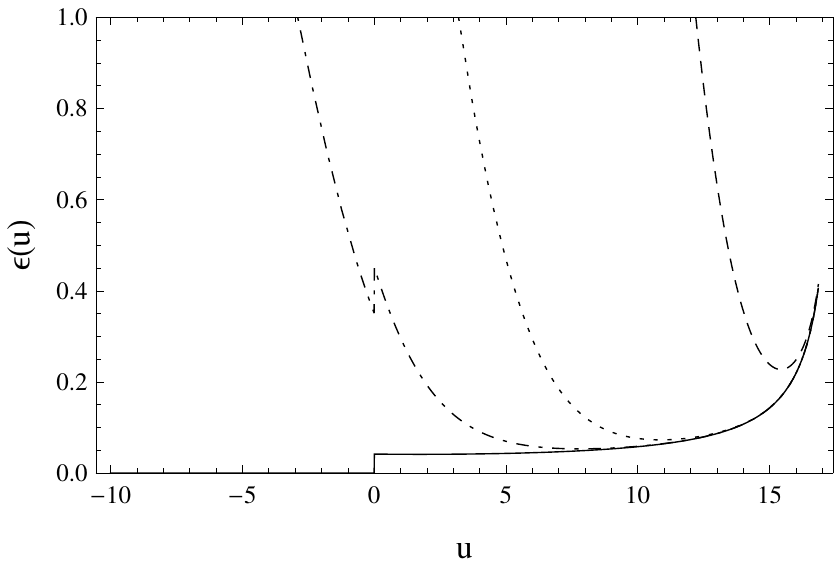}
  \caption{Adiabatic control function $\epsilon$ as a function of $u$ for $r_0 = 10 m$ and delays $\dtw =$ ($0$, $24 m$, $36 m$, $\infty$) [curves depicted respectively as (\dashed, \dotted, \chain, \full)]. We use $2m=1$ units and $u_0 = 0$.}
  \label{epsilon.times.radius.graph}
\end{figure}

\begin{figure}[ht]
	\centering
		\includegraphics{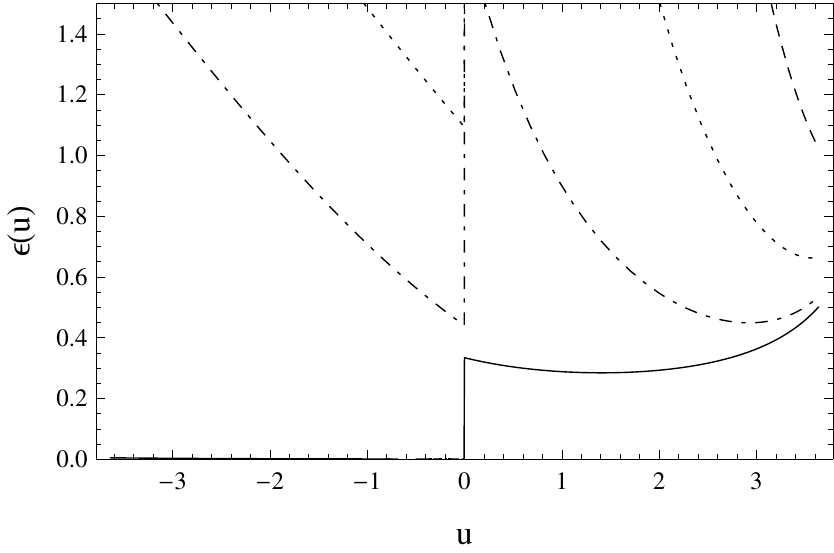}
  \caption{Adiabatic control function $\epsilon$ as a function of $u$ for $r_0 = 4 m$ and delays $\dtw =$ ($0$, $4 m$, $10 m$, $\infty$) [curves depicted respectively as (\dashed, \dotted, \chain, \full)]. We use $2m=1$ units and $u_0 = 0$.}
  \label{epsilon.times2.radius.graph}
\end{figure}

\begin{figure}[ht]
	\centering
		\includegraphics{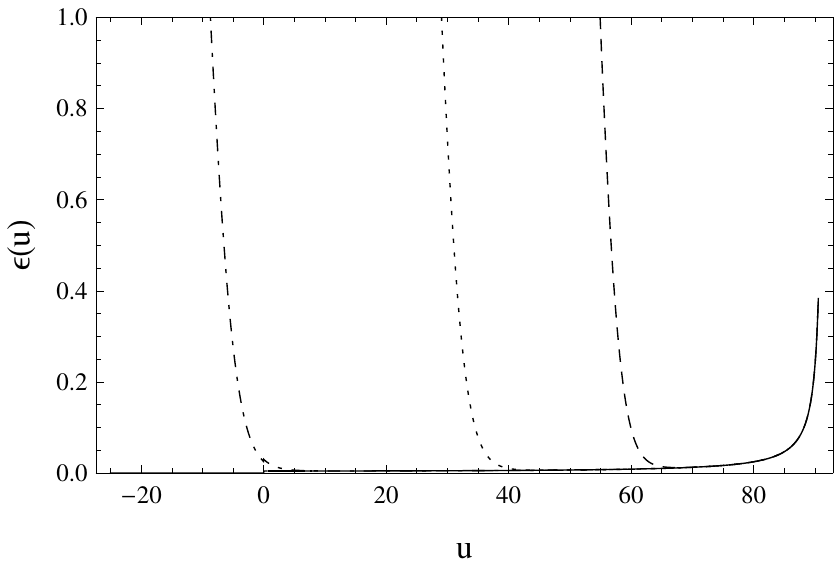}
  \caption{Adiabatic control function $\epsilon$ as a function of $u$ for $r_0 = 30 m$ and delays $\dtw =$ ($0$, $60 m$, $140 m$, $\infty$) [curves depicted respectively as (\dashed, \dotted, \chain, \full)]. We use $2m=1$ units and $u_0 = 0$.}
  \label{epsilon.times3.radius.graph}
\end{figure}

Finally, we will plot the value of $\epsilon (u)$ for the same cases we plotted 
$\kappa(u)$ in \sref{numerical.kappa.radius} 
(figures~\ref{epsilon.times.radius.graph}, \ref{epsilon.times2.radius.graph}, 
and~\ref{epsilon.times3.radius.graph}). As a consequence of the sudden change in 
$\kappa$ at the release time, these graphs should exhibit a Dirac-delta peak in there.
 This has been avoided in the figures for simplicity. Apart from that, graphs need 
few comments. They of course reproduce the limit values already found.
 As
 happened in \sref{freely.falling.infinity}, when plateaus appear   
$\epsilon$ 
becomes nearly zero, and so the radiation is perfectly thermal, with the 
temperature determined by $\kappa$. For earlier parts of the trajectory, $\epsilon$ 
diverges, but as we have already explained, this is only reflecting that the 
temperature there goes to zero.

\newpage

%---------------------------------------------------------------------
\section{Final increase of the effective temperature}
\label{Sec:interpretation}
%---------------------------------------------------------------------

At this stage, we have all the necessary ingredients to explain the, at first sight 
puzzling, final increase of $\kappa$ for freely-falling observers. For sufficiently long 
waiting times, we are sure the vacuum state is indistinguishable from Unruh state 
(see also next section). As standard lore in the literature, one can read that this state
 is a vacuum for observers freely falling at the horizon. However, we have here 
several of these observers and the value of $\kappa_{\rm HC}$ is non-zero for 
almost all of them: recall for instance the value of $\kappa_{\rm HC}$ for 
observers freely falling from infinity~\eref{kappa.HC.falling} and for observers 
freely falling from a finite radius~\eref{kappa.HC.radius}. The only exceptions are 
the observer closely following  the reference trajectory and the one 
with zero instantaneous radial velocity at the horizon.

Indeed, once Hawking radiation is switched on, there is only one effective temperature that does vanish at the horizon: the Unruh temperature 
$\kappa_{\rm Unruh}/(2 \pi)$~\eref{kappa.Unruh.radius}. As we explained, this is the 
$\kappa$ associated with an observer in free fall at a radial position $r$ and with an instantaneous 
zero radial velocity in there (we will call this observer an Unruh observer),
\begin{equation}
\kappa_{\rm Unruh} (r) = \fr{4m} \left(1 +\frac{2m}{r}\right) \left(1 -\frac{2m}{r}\right)^{1/2}.
\label{kappa.Unruh.radius.r}
\end{equation}

\begin{figure}[ht]
	\centering
		\includegraphics{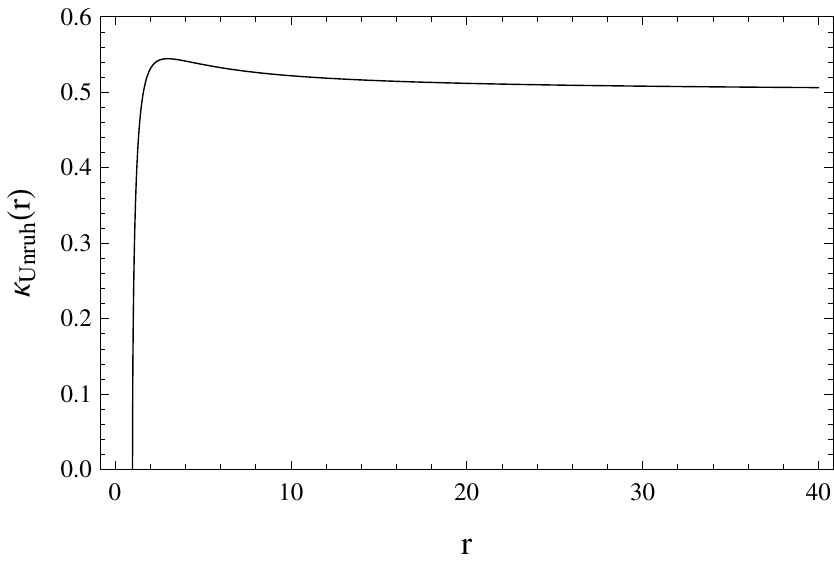}
  \caption{Temperature for the Unruh observer $\kappa_{\rm Unruh}$ as a function of $r$. We use $2m=1$ units.}
  \label{kappa.unruh.graph}
\end{figure}

Let us now make the reasonable hypothesis that all the physics concerning the behaviour of $\kappa(u)$ can be understood in terms of local properties of the observer's trajectory: its position, its velocity and its acceleration. 
We will see that this is indeed the case. Under this hypothesis, the only characteristic that distinguishes the Unruh observer from any of the other freely-falling observers is that the latter have non-zero radial velocity. 
The difference in radial velocities between the freely-falling observers and the Unruh observer implies that their respective radiation perceptions have to be related by a Doppler shift factor. 
In fact, the Unruh observer is such that in the horizon limit would travel with the very outgoing null ray. 
Therefore, to obtain $\kappa(u)$ for the freely-falling observers from $\kappa_{\rm Unruh}$, one has to extract from it a Doppler red-shift factor (going to zero at the horizon), or equivalently multiply it by a Doppler blue-shift factor (its inverse, diverging at the horizon). As we are going to show, this divergence counterbalance the zero value of $\kappa_{\rm Unruh}$ at the horizon yielding the appropriate finite result for the different $\kappa_{\rm HC}$'s.

Which is the precise form of this Doppler blue-shift factor? Our guess is
\begin{equation}
D_{r_0}(r):=\left(\deriv{r_{\rm l}}{t} - \deriv{r_{r_0}}{t}\right)^{1/2}  \left( \deriv{r_{\rm l}}{t} + \deriv{r_{r_0}}{t} \right)^{-1/2}.
\label{doppler}
\end{equation}
Here, $r_{\rm l}(t)$ represents the trajectory of a null ray going away from the horizon and $r_{r_0}(t)$ the freely-falling trajectory that starts with zero velocity at $r=r_0$. In addition, we have put a subscript $r_0$ in $D_{r_0}(r)$ to distinguish the different Doppler factors associated with freely-falling observers released at different radial distances $r_0$. The reason behind this guess is the following.

A physical Doppler factor has to compare the velocity of the massive objects with 
the velocity of light,
\begin{equation}
D= \sqrt{c - v \over c+v}.
\label{doppler.generic}
\end{equation}
However, note  that using $(t,r)$ coordinates the velocity of light is  different from 
unity; that is why we have  included ${\rmd r_{\rm l}}/{\rmd t}$. The
 Doppler factor~\eref{doppler} can be  alternatively written as
\begin{equation}
D_{r_0}(r)=\left(1 - \deriv{r_{r_0}}{r_{\rm l}}\right)^{1/2} \left( 1 + 
\deriv{r_{r_0}}{r_{\rm l}} \right)^{-1/2}.
\label{doppler.rel}
\end{equation}
In this way, one would compare the real local velocity of the observer with respect to the local velocity of light (which is always equal to one).

We can take ${\rmd r_{r_0}}/{\rmd t}$ from~\eref{dr.of.t.radius} and calculate ${\rmd r_{\rm l}} / {\rmd t}$ from the outgoing rays equation $t-r^* (r_{\rm l}) = \cte$,
\begin{equation}
\evat{\deriv{r_{\rm l}}{t}}{r_{\rm l}=r} = \deriv{r}{r^*}=\left(1-{2m \over r}\right).
\label{doppler.loc}
\end{equation}
After some manipulation, we finally find
\begin{eqnarray}
\fl D_{r_0}(r) = \left[ \left(1 -{2m \over r_0}\right)^{1/2} + \sqrt{2m \over r}\left(1 -{r \over r_0}\right)^{1/2} \right]^{1/2}
\nonumber\\
\times \left[ \left(1 -{2m \over r_0}\right)^{1/2} - \sqrt{2m \over r}\left(1 -{r \over r_0}\right)^{1/2}
\right]^{-1/2}.
\label{doppler.expression}
\end{eqnarray}

If we compute now 
\begin{equation}
\kappa_{\rm Unruh}(r) D_{r_0}(r),
\label{kappa.final.from.doppler}
\end{equation}
and take the limit $r \to 2m$, we can check that one obtains precisely~\eref{kappa.HC.Unruh.radius}. So, as we advanced, it is really the extraction of a huge Doppler red-shift factor the reason why the freely-falling observers perceived some effective temperature even at horizon crossing. In other words, for long enough waiting times, a generic freely-falling observer sees a small but non-zero $\kappa$ when crossing the horizon. 
Only a hypothetical observer with zero radial velocity at the horizon (the Unruh observer there) would see a zero $\kappa$, but because of the huge red-shift factor associated to its instantaneous trajectory (this observer is instantaneously travelling with the null ray). 

The final increase in $\kappa$ always present for freely-falling observers is caused by
 the Doppler shift factor. Take the case of an observer freely falling from 
infinity. 
As we explained, the perceived radiation starts at zero temperature. Then, if one waits long-enough, a plateau region is formed with an almost constant $\kappa$. 
Finally, when the observer approaches the horizon, $\kappa$ increases up to the value $1/m$, which corresponds to four times the standard Hawking temperature. The value of $\kappa$ from the appearance of the plateau on is due to the \emph{competition} between Unruh's switching off of the radiation emission when going towards the horizon, which surprisingly passes through an intermediate increasing phase (see~\fref{kappa.unruh.graph}), and the multiplication by a Doppler blue-shift factor due to the velocity of the freely-falling observer with respect to the Unruh observer. 
The final result of this competition is the exact numerical computations we have shown throughout the paper (considering always sufficiently long  waiting times).

\begin{figure}[ht]
	\centering
		\includegraphics{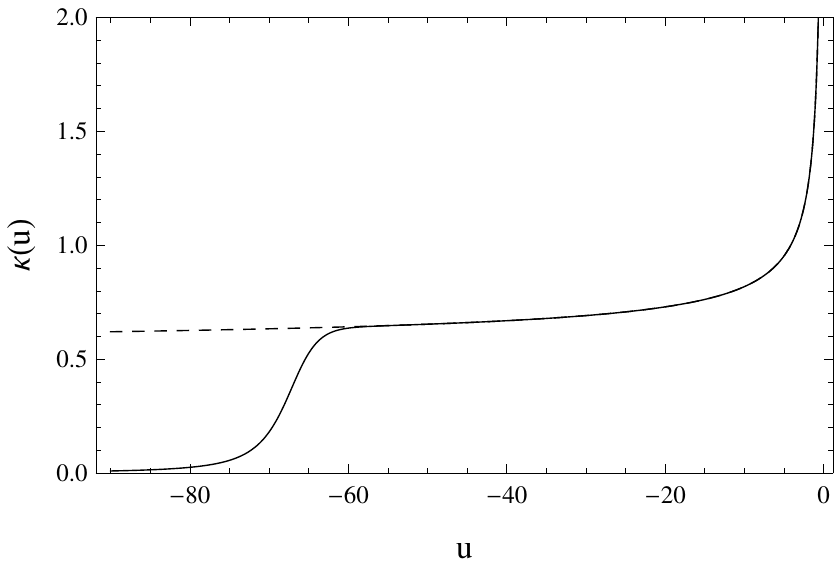}
  \caption{Temperature as a function of $u$ for a freely-falling observer from infinity with $\dt = 200 m$ using exact expression (curve depicted \full) and using $\left( \kappa_{\rm Unruh} D_{r_0 \to \infty} \right) (u)$ approximation (curve depicted \dashed). We use $2m=1$ units and $\uH = 0$.}
  \label{plateau.falling.graph}
\end{figure}

For instance, let us plot  
\begin{equation}
\kappa_{\rm Unruh}(r(u))D_{r_0 \to \infty}(r(u))
\label{kappa.plateau}
\end{equation}
against the exact numerical $\kappa(u)$ of \sref{freely.falling.infinity}. Here $r(u)$ is the position of the observer falling from infinity as a function of the label $u$. From~\eref{r.of.tau} one obtains $r(u)$
\begin{equation}
r(u) = 2m \left[ {3 \over 4m}(\uH-u) + 1 \right]^{2/3}.
\label{r.of.u.falling}
\end{equation}
In \fref{plateau.falling.graph}, we compare this approximation with the exact result for $\kappa(u)$ for a long waiting time $\dt = 200 m$. We can see that the approximation is perfect from the starting of the plateau till horizon crossing.
Obviously, the previous approximation fails to capture the switching on of Hawking radiation.

Actually, this perfect fit for long waiting times can also be checked analytically, and not only for observers falling from infinity, but for the generic case of freely-falling observers from any radius. For instance, one just has to approximate the $\kappa(U,r)$ in~\eref{kappa.radius} for $\dtw \to \infty$ (this is equivalent to assuming $U \simeq U_H$), which yields 
\begin{eqnarray}
\fl \kappa(r) = \fr{4m} \left\{\frac{r}{2m}+1 +\frac{r}{m} \left[ \left(1-\frac{2m}{r_0}\right)^{1/2} \sqrt{\frac{2m}{r}}\left(1-\frac{r}{r_0}\right)^{1/2} - \frac{2m}{r_0} \right]\right\}
\nonumber\\
 \times \left\{\left[ \left(1-\frac{2m}{r_0}\right)^{1/2} + \sqrt{\frac{2m}{r}}\left(1-\frac{r}{r_0}\right)^{1/2} \right] \left(\frac{r}{2m}-1\right)\right\}^{-1}
\nonumber\\
\times \left[1 -\left({2m \over r}\right)^2 \right],
\label{kappa.radius.dt.infinity}
\end{eqnarray}
and then see that it is precisely equal to
\begin{eqnarray}
\fl \kappa_{\rm Unruh}(r) D_{r_0}(r) =  \fr{4m} \left(1 +\frac{2m}{r}\right) \left(1 -\frac{2m}{r}\right)^{1/2}
\nonumber\\
 \times
\left[ \left(1 -{2m \over r_0}\right)^{1/2} + \sqrt{2m \over r}\left(1 -{r \over r_0}\right)^{1/2} \right]^{1/2}
\nonumber\\
\times \left[ \left(1 -{2m \over r_0}\right)^{1/2} - \sqrt{2m \over r}\left(1 -{r \over r_0}\right)^{1/2}
\right]^{-1/2}.
\label{kappa.radius.dt.infinity2}
\end{eqnarray}

With hindsight, we could have used this equality to obtain all the previous partial 
results. But we can go now even further: if we put $\kappa_{\rm released} 
(r,\dtw)$ instead of $\kappa_{\rm Unruh} (r)$ in~\eref{kappa.radius.dt.infinity2}, 
then it is possible to see that we reproduce the entire formula for the effective temperature 
in~\eref{kappa.radius}. This is because between two freely-falling observers at the 
same position the only difference is a Doppler shift, and this fact has nothing to do 
with being in the Unruh state or not. We checked this first for the Unruh state because we wanted to explain the appearance of the final peak in $\kappa$, a peak whose appearance
was in apparent contradiction with the vanishing value we expected for
that 
state.

%---------------------------------------------------------
\section{Comparison with the Unruh state}
\label{Sec:Unruh}
%---------------------------------------------------------

One might worry that some of the results shown in this paper related with the late-time behaviour of the system, are peculiarities of an unusual choice of the vacuum state. However, this is not the case. In this section, we will briefly show that at late times this state is really equivalent to the Unruh vacuum state. Thus, all the properties found in the previous sections that applies for late times, are immediately present also in the 
Unruh state.

The equivalence between the states for late times can be directly seen 
from~\eref{ub.of.U}. In that expression, late times means just $U \simeq \UH$. So,
 as we did in~\eref{u.of.U.static.future}, we can reverse the approximate relation 
to find
\begin{eqnarray}
\fl U \approx \UH -\frac{4m}{3} \left\{ \left[ \exp \left(\fr{4m} \left(\UH - 6m \right) \right)
\rme^{-\ub/(4m)} +1 \right]^3 - 1 \right\} 
\nonumber\\
\approx \UH - 4m \exp \left[\fr{4m} \left(\UH - 6m \right) \right] \rme^{-\ub/(4m)}.
\label{U.of.ub.future}
\end{eqnarray}
But this is precisely the relation which defines the coordinate $U$ that appears in the 
Unruh modes~\cite{unruh} (apart from the irrelevant origin of the 
coordinates). Thus, if we are using this coordinate to define the vacuum state, we 
are in fact choosing the Unruh state for late times.

Nonetheless, just as an example, let us perform a specific calculation 
entirely based in the Unruh state choice. For instance, let us calculate the case of a freely-falling 
observer from infinity. With this aim, consider a redefinition of $U$ using the usual and 
simple form
\begin{equation}
U := - 4m \rme^{-\ub/(4m)},
\label{U.of.ub.Unruh}
\end{equation}
and take~\eref{ub.of.u.falling} as the definition of $u$. That is, $U$ defines the vacuum and is the coordinate that appears in the Unruh modes; and $u$ is the proper time that a freely-falling observer from infinity uses to label the rays. Replacing~\eref{ub.of.u.falling} in~\eref{U.of.ub.Unruh} we obtain a relation $U(u)$, from which we can directly compute $\kappa(u)$. The result is
\begin{equation}
\fl \kappa(u) = \fr{4m} \left\{ \left[ \frac{3}{4m} \left(\uH - u \right) + 1 \right]^{-1/3} + 2 \left[ \frac{3}{4m} \left(\uH - u \right) + 1 \right]^{-2/3} + 1 \right\}.
\label{kappa.Unruh.state}
\end{equation}

\begin{figure}[ht]
	\centering
		\includegraphics{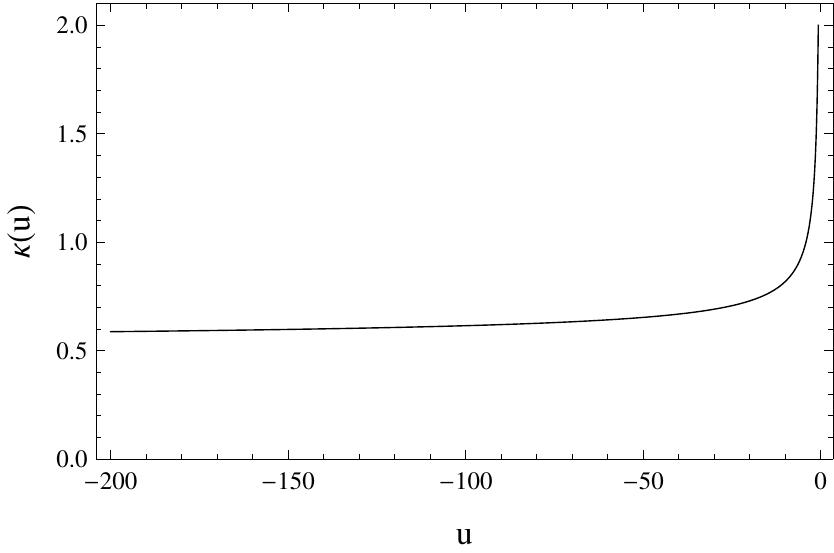}
  \caption{Temperature as a function of $u$ for a freely-falling observer from infinity 
in the Unruh state. We use $2m=1$ units and $\uH = 0$.}
  \label{unruh.state.graph}
\end{figure}

From here one can check the limits
\begin{equation}
\kappa (u \to -\infty) = \fr{4m} \quad {\rm and} \quad \kappa (u \to \uH) = \fr{m}
\end{equation}
with the naked eye. The first limit is the usual Hawking radiation temperature, which in the Unruh state is always present, even at the asymptotic past. It is as if we extended the plateau to the $u \to -\infty$ limit (see \fref{unruh.state.graph}). The second limit is equal to the one obtained at the horizon 
crossing for long enough delays in \sref{freely.falling.infinity} 
[see~\eref{kappa.HC.falling}], and also for far enough releasing radius in 
\sref{freely.falling.radius} [see~\eref{kappa.HC.infinity.radius}]. Thus, the final 
peak in the temperature is not something coming from an eccentric selection of the 
state: it is a characteristic of the Unruh state.
%

%---------------------------------------------------------
\section{Summary and conclusions}
%---------------------------------------------------------

In this paper we have considered a quantum scalar field over a Schwarzschild black 
hole geometry. As a first step, we have set up the quantum scalar field in a specific 
vacuum state that interpolates between Boulware vacuum state at early times and 
Unruh vacuum state at late times. In other words, we have chosen a non-stationary 
vacuum state so that it simulates the physics of a dynamical collapse scenario. In 
this scenario, before the black hole forms there is no radiation at infinity 
but,
 after its formation, observers at infinity start receiving a Hawking flux of radiation: 
the black hole Hawking emission switches on at some particular ignition time (of 
course, it is not an instantaneous event). This vacuum state has been 
fixed 
by requiring that a reference observer freely falling from infinity detects 
no 
radiation.

As a second step, we have analysed the radiation associated with this vacuum state 
in terms of how it is perceived by different observers: static observers at
 a fixed radius; freely-falling observers from infinity, with different time delays with 
respect to the reference observer; and observers maintained at a fix 
radius for some time, and then, after being surpassed by the reference observer in 
its way in, released to fall freely towards the horizon at different releasing times.

The method used follows the recent analysis in~\cite{Barcelo:2010pj, Barcelo:2010xk}. For each observer we define a function $\kappa$ varying along the observer's trajectory. We also define an adiabatic control function $\epsilon$ as $\epsilon=|\hfrac{\rmd \kappa}{\rmd u}|/\kappa^2$. In the portions of the trajectory in which $\kappa$ stays almost constant ($\epsilon \ll 1$) one can be sure that the observer detects thermal radiation at a slowly varying temperature $k_{\rm B}T =\kappa/2\pi$. When this is not exactly the case, the radiation perceived is non-thermal, but still, we take the value of $\kappa$ as an estimate of the amount of perceived radiation and, somewhat abusing of the language, we talk about a time varying effective temperature.

\begin{itemize}

\item
{\it Static observers at fixed radius.}
We show that they perceived an effective temperature that starts from
zero, at early times, and after a transient period, stabilizes at Hawking's temperature
 $1/(8\pi m)$ multiplied by the gravitational blue-shift factor associated with the 
particular radial position of the observer. This description clearly reflects the 
\emph{ignition} associated with the formation of a black hole we were seeking for. 
We also see that the transient period is quite sort: $0.07 \sci{-3}{s}$ for a collapsing neutron star with 1.5 times the mass of the Sun and a radius of $12~\mathrm{km}$, when detected from far enough positions. Observers closer to the 
horizon will perceive this process with a slower pace due to the 
gravitational slow down of their clocks.

\item
{\it Observers freely falling from infinity.}
When their delay with respect to the reference trajectory used to fix the vacuum 
state is small, they perceive almost no radiation at all. The physical 
picture is the following: If an observer sitting at the very 
surface of a 
star that is collapsing in free fall to form a black hole  
perceived vacuum at the beginning of the collapse, he would perceive vacuum 
during his entire collapsing trajectory (this is reasonable, as he has not perceived any
 gravitational field).  

On the contrary, when the delay is long enough, the observer has time enough to 
see that the black hole is emitting Hawking radiation. He first sees the switching on of
 this radiation. Then, during some time he perceives an almost constant temperature
 radiation passing by. The temperature at this plateau is essentially the result of 
multiplying Hawking's temperature by a Doppler shift factor associated with the 
velocity of the observer. In this region the adiabatic condition is perfectly satisfied, so
 here to describe the perceived radiation as having a temperature is strictly correct. 
Finally, in the last stages of his approach to the horizon, and surprisingly at first sight,
 the effective temperature rises reaching exactly four times Hawking's 
temperature. The adiabatic condition is only mildly violated in these last stages, so 
although the perceived radiation is not exactly Planckian, it is not hugely different 
from Planckian either. 

\item
{\it Observers freely falling from a radius.}
These observers are standing still at a fix radial distance $r_0$ from the 
horizon (supporting themselves against the gravitational pull with some rockets) till
 a release time (when the rockets are switched off), which happens 
some waiting time after the reference observer, in free fall, has 
overtaken them. If this waiting time is long enough, the observers will 
see Hawking radiation appear and settle down to a constant value, 
right in the same way as described above for static observers. Once the observer 
is released, the value of the effective temperature undergoes a jump 
with a value equal to the surface gravity associated with position $r_0$ times the 
gravitational blue-shift factor associated with this same position. We 
calculated the value of the effective temperature after the release and   
called it $k_{\rm B} T_{\rm released}(r) = 
\kappa_{\rm released}(r)/ (2\pi) $. From the release time on, the 
effective temperature starts to rise, first slowly and later quicker, so as to reach a 
finite value at horizon crossing. The precise behaviour of the effective temperature 
depends on the release radius, and so does the asymptotic value of it, which 
becomes $4T_{\rm H}$ when $r_0 \to \infty$.  

\end{itemize}

Once Hawking radiation is switched on, the value of $k_{\rm B} T_{\rm Unruh}(r) = 
\kappa_{\rm Unruh}(r)/ (2\pi) $, that is, the effective temperature as perceived by an
 observer freely falling at $r$ and with a vanishing instantaneous radial
 speed, interpolates from $T_{\rm H}$ when $r \to \infty$ to zero when $r 
\to 2m$. This is consistent with the idea that the Unruh state is vacuum for freely 
falling observers at the horizon. However, in general one has to be careful with this 
assertion. We have shown that after Hawking radiation is switched on, 
generic freely-falling observers at the horizon \emph{do not perceive 
vacuum, but a finite effective temperature higher than their initial Unruh effective 
temperature}. The reason is that the Unruh observer experiences a Doppler red-shift
 factor with respect to a generic observer (due to their velocity difference), or in 
other words, a generic observer experiences a Doppler blue-shift factor with the 
respect to the Unruh observer. This Doppler blue-shift factor diverges at the horizon 
in such a way that, when multiplied by the vanishing value of Unruh's 
temperature at the horizon, results in a finite temperature for a generic observer.   
In fact, we have proved the following completely general result: for any 
freely-falling observer the exact effective temperature $T (r)$ 
perceived at position $r$ is precisely the product of the 
temperature when released there $T_{\rm released}(r)$ by the Doppler blue-shift
factor associated with the radial velocity of the observer when passing 
through that position.

Regarding the observability of Hawking radiation, in pure theoretical terms the discussed results lead to the idea that the closer to the horizon one supports oneself against gravity, the better the chances of detecting its radiation, due to the gravitational blue-shift factor. However, in practical astrophysical scenarios the tiny Hawking temperature of a stellar mass black hole, about a few nano-Kelvins, will be masked by the $\sim 3$ Kelvin of the cosmic microwave background. This radiation would also experience a gravitational blue shift if detected by observers close to the horizon, so that going closer to the horizon would not improve in principle the chances of detection. Another matter is the detection of Hawking radiation in analogue models of General Relativity. There exist quantum systems, as Bose-Einstein condensates, in which Hawking temperature is not that much masked by the environment temperature. In those situations one could think that analysing the presence of radiation close to the horizon would improve the detection chances, but this is not that clear. These systems exhibit modified dispersion relations so that when blue shifting the modes, they would enter into the non-relativistic regime, behaving very differently to what has been found and discussed in this paper. In addition, the size of the entire system in the lab will be so small that to control the physics localized close to horizon, a tiny fraction of the total size of the system, seems implausible. Other ways of detection that could prove better are the so called correlation measurements between outgoing and ingoing partners of the Hawking radiation~\cite{correlations, correlations2}.

Let us finish the paper by mentioning that the results presented here will also constitute an 
adequate preparatory background material for a future work dealing with analyses of 
buoyancy over black hole horizons.

%---------------------------------------------------------
\ack
%---------------------------------------------------------

The authors want to thank S. Finazzi, S. Liberati, S. Sonego and M. Visser 
for enlightening comments during the elaboration of this paper.
L. C. B. and C. B. have been supported by the Spanish MICINN through the project 
FIS2008-06078-C03-01 and by the Junta de Andaluc\'{\i}a through the projects 
FQM2288 and FQM219. L. J. G. has been financially supported by the 
Spanish MCINN through the project FIS2008-06078-C03-03 and the Consolider-Ingenio 
2010 Program CPAN (CSD2007-00042).

%---------------------------------------------------------

%---------------------------------------------------------
\end{document}